\documentclass[a4paper,fleqn]{cas-sc}

\usepackage[authoryear]{natbib}
\usepackage{graphicx}
\usepackage{amsmath}
\usepackage{amssymb}
\usepackage{amsthm}
\usepackage{amsfonts}
\usepackage{bm}
\usepackage{algorithm}
\usepackage{algpseudocode}
\usepackage{float}
\usepackage{multirow}
\usepackage{longtable}

\newtheorem{theorem}{Theorem}
\newtheorem{corollary}{Corollary}[theorem]
\newtheorem{remark}{Remark}
\newtheorem{definition}{Definition}
\newtheorem{lemma}{Lemma}

\def\tsc#1{\csdef{#1}{\textsc{\lowercase{#1}}\xspace}}
\tsc{WGM}
\tsc{QE}
\tsc{EP}
\tsc{PMS}
\tsc{BEC}
\tsc{DE}

\begin{document}
\let\WriteBookmarks\relax
\def\floatpagepagefraction{1}
\def\textpagefraction{.001}

\shorttitle{Eco-MP}

\shortauthors{Tan}

\title [mode = title]{\textbf{Eco-MP: Ecological Max-Pressure Traffic Signal Control with Ecological Driving}}

\author[1]{Haoran Wang}[orcid=]
\ead{wang_haoran@tongji.edu.cn}
\fnmark[1]

\author[1]{Shuoyuan Li}[orcid=]
\ead{shuoyuanli@tongji.edu.cn}
\fnmark[1]

\author[2]{Chaopeng Tan}[orcid=0000-0003-4737-5304]
\ead{chaopeng.tan@tu-dresden.de}
\credit{Conceptualization of this study, Methodology, Writing}
\cormark[1]

\author[3]{Hao Liu}[orcid=]
\ead{hao.liu1@maine.edu}

\author[1]{Keshuang Tang}[orcid=]
\ead{tang@tongji.edu.cn}

\author[1]{Jia Hu}[orcid=]
\ead{hujia@tongji.edu.cn}

\cortext[1]{Corresponding author}
\fntext[1]{These authors contributed equally to this work.}

\affiliation[1]{organization={The Key Laboratory of Road and Traffic Engineering of the Ministry of Education, College of Transportation Engineering, Tongji University},
    addressline={No. 4800, Cao'an Road},
    city={Shanghai},
    postcode={201804},
    country={China}}

\affiliation[2]{organization={Chair of Traffic Process Automation, Technische Universität Dresden},
    addressline={Hettnerstraße 3}, 
    city={Dresden},
    postcode={01069}, 
    country={Germany}}

\affiliation[3]{organization={Department of Civil and Environmental Engineering, University of Maine},
    addressline={168 College Ave}, 
    city={Orono},
    postcode={ME 04469}, 
    country={USA}}

\begin{abstract}
Max-pressure (MP) traffic signal control provides a decentralized mechanism for urban networks, but conventional pressure states do not directly capture the propulsion burden generated by signal-induced stop-and-go traffic. Eco-driving, by contrast, typically relies on prespecified signal information and optimizes vehicle trajectories separately from the feedback among vehicle motion, traffic states, and signal decisions.
This paper develops Ecological Max-Pressure (Eco-MP), a closed-loop framework that integrates ecological MP signal control with MP-adaptive ecological driving. At the signal layer, connected-vehicle observations are aggregated into a common movement state based on cumulative positive propulsion work, and the resulting upstream--downstream ecological pressure determines phase selection in mixed petrol and electric traffic. At the vehicle layer, a rolling signal interface translates the evolving Eco-MP process into near-term signal information, while each connected and automated vehicle solves a powertrain-independent finite-horizon trajectory problem subject to vehicle-dynamic, car-following, and signal-compliance constraints to reduce unnecessary stopping and acceleration effort.
Under full-state information, a finite-storage Lyapunov analysis establishes strong stability for admissible demands satisfying a strict supportability condition and extends the result to the coupled signal--vehicle system. When ecological driving enlarges the admissible-flow sets, the coupled admissible demand region contains the signal-only region.
Microscopic experiments on an Amsterdam arterial corridor show that, relative to queue-based MP, Eco-MP reduces mean positive propulsion work by 5.2\%, vehicle delay by 20.8\%, and stopping time by 39.0\%. With only 15\% of vehicles directly controlled, ecological driving further reduces positive propulsion work by 4.2\% and stopping time by 19.6\%. Performance improves as connected-vehicle penetration increases and varies only modestly across the tested petrol-vehicle shares.
\end{abstract}

\begin{keywords}
traffic signal control \sep max-pressure control \sep eco-driving \sep model predictive control \sep connected and automated vehicles
\end{keywords}

\maketitle

\section{Introduction}

Urban transportation systems are increasingly expected to provide efficient mobility while reducing energy use and environmental impacts. Signalized intersections are particularly important because vehicles repeatedly decelerate, wait, restart, and accelerate around signal transitions, creating concentrated stop-and-go propulsion demand and emissions \citep{pandian2009intersection,choudhary2016urban}. Under congestion, these cycles can persist and propagate upstream, further increasing the energy burden of urban traffic \citep{grote2016congestion}. Ecological performance at an intersection is therefore shaped jointly by two control processes: signal service determines when a movement can discharge, whereas the realized vehicle trajectory determines the propulsion effort required before and after that service. Network-level signal decisions can consequently affect not only mobility but also traffic-related energy use and emissions \citep{osorio2015emissions}. Improving ecological performance at signalized intersections thus requires coordinated consideration of signal control and vehicle motion.

Max-pressure (MP) control provides a natural foundation for the signal-control side of this problem because it makes decentralized phase decisions from local upstream and downstream states and provides queue-stability guarantees under admissible demand \citep{varaiya2013max}. A key feature of MP is that its behavior can be shaped through the state used to construct pressure. Capacity-aware MP accounts for downstream storage when evaluating service \citep{gregoire2014capacity}, while position-weighted MP incorporates the spatial distribution of vehicles within links \citep{li2019position}. Delay-based formulations use temporal congestion burden to modify phase priorities \citep{wu2018delay,liu2022novel,liu2023total}, and travel-time-based MP further exploits trajectory information observed from connected vehicles \citep{mercader2020max,zoabi2025experimental}. Connected-vehicle-based MP has also been developed for heterogeneous and partial connectivity, where movement states and turning information must be inferred from incomplete observations \citep{tan2026cv}; related work has considered privacy-preserving heterogeneous data fusion for MP state construction \citep{tan2026private}. Beyond traffic-state representation, the MP framework has been extended to phase-order constraints \citep{levin2020cyclical}, multimodal and transit priority \citep{xu2022integrating,xu2024ped,liu2024pedqueue,tan2027transit}, and broader coordination problems such as perimeter regulation and route guidance \citep{ahmed2024occ,le2017utility,ahmed2025c}. Together, these developments show that the pressure state and associated control structure can be adapted to different operational objectives while retaining decentralized phase selection.

Existing MP states, however, remain primarily congestion- or service-oriented. Queue length represents how much traffic awaits service, delay and travel time represent how long traffic has been impeded, and position or storage information describes the spatial distribution of congestion. None of these quantities directly records the propulsion burden accumulated as vehicles brake, idle, restart, and accelerate through a signalized movement. Two movements with similar queue lengths or travel times can therefore have different ecological consequences because their vehicles have experienced different speed and acceleration histories. This distinction becomes more important in mixed-powertrain traffic, where petrol and electric vehicles translate the same motion pattern into energy use through different powertrain mechanisms. A signal controller based directly on fuel or electricity consumption would therefore require powertrain-specific state quantities. A more general formulation should preserve the upstream--downstream pressure structure of MP while using an ecological state that can be applied consistently across powertrain types.

A parallel literature has developed eco-driving and signal--vehicle cooperative control at the vehicle level. With usable signal phase and timing information, an approaching vehicle can anticipate a red interval, decelerate earlier, avoid an unnecessarily fast approach, and exploit an available green window more effectively \citep{huang2018ecodrivingreview,li2024ecodrivingreview}. Signal-aware trajectory optimization has been shown to reduce energy use for individual vehicles \citep{li2018evsignal}, and cooperative formulations have extended this idea to mixed automated and human-driven platoons \citep{zhao2018platoon}. Integrated signal--trajectory methods jointly optimize signal timing and vehicle motion through mixed-integer, single-layer, or hierarchical formulations \citep{yu2018integrated,liu2022singlelayer,zou2024integrated}. Other studies have addressed multiple intersections \citep{dong2022predictive}, uncertain traffic and signal conditions \citep{lu2024uncertainty}, and lane-changing disturbances \citep{hu2024urban}. Vehicle advisory has also begun to be combined with MP-type signal control in transit applications; for example, dwelling and speed advisory have been integrated with max-pressure transit signal priority \citep{tan2026dwelling}. Collectively, these studies show that anticipatory vehicle motion can reduce unnecessary stopping and acceleration when suitable signal information is available.

The two control streams are not yet naturally coupled in a feedback-driven MP setting. Many eco-driving formulations treat future signal information as fixed or externally provided \citep{li2024ecodrivingreview,zhao2018platoon,dong2022predictive}. Joint signal--trajectory optimization closes this loop by optimizing signal timing and vehicle motion together \citep{yu2018integrated,liu2022singlelayer,zou2024integrated}, but it follows a different architecture from decentralized MP, in which future phases are generated online from traffic states that continue to evolve. A trajectory planned against a signal sequence that is inconsistent with the active MP process can miss a usable green interval or cause unnecessary stopping. More importantly, vehicle control is itself part of the network dynamics: changes in speed and acceleration alter both traffic evolution and the ecological state observed by subsequent signal decisions. Integrating eco-driving with MP therefore requires a rolling interface that translates the evolving MP process into usable near-term signal information, together with a coupled analysis that remains consistent with the trajectory-modified traffic process.

This paper develops Ecological Max-Pressure (Eco-MP), a closed-loop framework for ecological signal and vehicle control in mixed-powertrain traffic. Connected vehicles (CVs) provide motion states and intended turning movements for ecological-state and turning-proportion estimation. Connected and automated vehicles (CAVs) form the directly controlled subset and additionally execute MP-adaptive ecological-driving commands. At the signal layer, vehicle motion is mapped to a common state based on cumulative positive propulsion work after vehicles enter the detected queue region; this state is then used to construct upstream--downstream ecological pressure. At the vehicle layer, the current Eco-MP process provides a rolling near-term signal prediction, and each CAV independently solves a powertrain-independent finite-horizon quadratic program subject to vehicle-dynamic, car-following, and signal constraints. The executed trajectories feed back into the evolving traffic and ecological states used by subsequent signal decisions.

The main contributions are as follows:
\begin{enumerate}
    \item We formulate an ecological MP signal controller using cumulative positive propulsion work as the movement state. Queue-experience memory retains each vehicle's contribution from its first entry into the detected queue region until stop-line discharge, and the resulting state is embedded in the standard upstream--downstream pressure structure. The same positive propulsion work is used for petrol vehicles (PVs) and electric vehicles (EVs), allowing a common ecological pressure and signal objective to serve a mixed-powertrain traffic stream.

    \item We develop an MP-adaptive ecological-driving layer for CAVs. A rolling near-term signal description is generated from the active Eco-MP process and incorporated into an independent receding-horizon trajectory problem for each controlled vehicle. The same kinematic optimization structure is used across powertrains, while fuel and electricity consumption are evaluated separately for the resulting trajectories. The vehicle controller can therefore respond directly to feedback-driven signal decisions while preserving the distributed structure of Eco-MP.

    \item We establish stability properties for both the signal-only and coupled systems using a finite-storage Lyapunov function matched to the cumulative ecological state. Under full-state information, Eco-MP and the coupled signal--vehicle system are strongly stable for admissible demands satisfying the stated strict supportability condition. A comparison of admissible demand regions further shows that the coupled region contains the signal-only region whenever ecological driving enlarges the corresponding admissible-flow sets.
\end{enumerate}

The remainder of the paper is organized as follows. Section~\ref{sec:pre} introduces the traffic-flow and vehicle-motion models. Section~\ref{sec:framework} presents Eco-MP and MP-adaptive ecological driving. Section~\ref{sec:stability} establishes stability results for the signal-only and coupled systems. Section~\ref{sec:evaluation} evaluates ecological, mobility, discharge, and network performance. Section~\ref{sec:conclusion} concludes the paper and outlines directions for future research.

\section{Preliminaries}\label{sec:pre}

This section defines the traffic-flow and vehicle-motion quantities used throughout Eco-MP. The network model describes movement-level conservation, signal-dependent realizable discharge, and turning-based propagation. The vehicle model maps longitudinal motion to positive propulsion work, which is later used to construct the ecological movement state.

\subsection{Network definition and traffic-flow dynamics}\label{sec:network_def}

Eco-MP uses a movement-based representation so that the state awaiting service can be compared with the state that discharged vehicles are expected to encounter downstream. The formulation follows the continuous-link representation used in position-weighted and connected-vehicle-based MP control \citep{li2019position,tan2026cv}.

Consider a signalized road network with a set of real signalized intersections $\mathcal N$ and a set of fictitious source nodes $\mathcal F$. Fictitious nodes represent exogenous demand entering the modeled network and provide point queues in which unmet boundary demand can accumulate. For each node $n\in\mathcal N\cup\mathcal F$, let $\mathcal M_n$ denote its movement set. A movement is indexed by $m=(i,o)$, where $i$ and $o$ denote its incoming and outgoing links, respectively, and $L_i$ denotes the length of incoming link $i$. The movement sets associated with real and fictitious nodes are $\mathcal M_{\mathcal N}\triangleq\bigcup_{n\in\mathcal N}\mathcal M_n$ and $\mathcal M_{\mathcal F}\triangleq\bigcup_{n\in\mathcal F}\mathcal M_n$. Real links have finite length and storage capacity, whereas a fictitious source link has no physical length and is represented as a point queue with effectively unbounded storage.

Let $\bm s(t)$ collect the movement-level signal indicators $s_m(t)\in\{0,1\}$, where $s_m(t)=1$ indicates that movement $m$ is served at time $t$. For each real intersection $n$, let $\mathcal P_n$ denote its feasible phase set, and let $s_m^p\in\{0,1\}$ indicate whether movement $m\in\mathcal M_n$ is served by phase $p\in\mathcal P_n$. The feasible set $\mathbb S$ contains the network signal states formed by compatible local phases at all real intersections. The sole movement associated with each fictitious source node is treated as continuously available for admission into the physical network.

For a real movement $m=(i,o)\in\mathcal M_{\mathcal N}$, let $\rho_m(x,t)$ and $q_m(x,t)$ denote the density and flow at position $x\in(0,L_i)$ on incoming link $i$ for vehicles assigned to outgoing link $o$. This movement-specific density preserves turning information in the traffic state. Let $q_m^{\mathrm{in}}$ and $q_m^{\mathrm{out}}$ denote the boundary inflow and outflow, respectively, and let $s_m^{\mathrm{in}}(t)$ and $s_m^{\mathrm{out}}(t)$ denote the signal indicators governing the corresponding upstream and downstream transfers. For a fictitious movement $m\in\mathcal M_{\mathcal F}$, $\rho_m(t)$ denotes its point-queue state. Its conservation equation is
\begin{equation}
\frac{d\rho_m(t)}{dt}
=
\lambda_m(t)
-
q_m^{\mathrm{out}}\!\left(s_m^{\mathrm{out}}(t)\right),
\qquad
m\in\mathcal M_{\mathcal F},
\label{eq:fictitious_dyn}
\end{equation}
where $\lambda_m(t)$ is the exogenous arrival rate and $q_m^{\mathrm{out}}$ is the rate admitted from the source queue into the physical network.

For a real movement, conservation along the incoming link is described by
\begin{equation}
\frac{\partial\rho_m(x,t)}{\partial t}
=
-\frac{\partial q_m(x,t)}{\partial x},
\qquad
x\in(0,L_i),\quad m\in\mathcal M_{\mathcal N},
\label{eq:continuity_pde}
\end{equation}
with boundary conditions
\begin{align}
q_m(0,t)
&=
q_m^{\mathrm{in}}\!\left(s_m^{\mathrm{in}}(t)\right),
\label{eq:upstream_boundary}\\
q_m(L_i,t)
&=
q_m^{\mathrm{out}}\!\left(s_m^{\mathrm{out}}(t)\right).
\label{eq:downstream_boundary}
\end{align}
Integrating Eq.~\eqref{eq:continuity_pde} over the link and applying Eqs.~\eqref{eq:upstream_boundary}--\eqref{eq:downstream_boundary} yields the movement-count balance between upstream inflow and downstream discharge used later in the stability analysis.

The signal action determines how much flow can actually discharge from a movement. Let $c_m>0$ denote the saturation flow rate of movement $m$, and let $\mu_m(t)\geq0$ denote its effective serviceable demand, determined jointly by available upstream traffic and downstream receiving capacity. For a candidate signal state $\bm s$, the realizable discharge is
\begin{equation}
d_m(\bm s,t)
=
\min\!\left\{c_m s_m,\,\mu_m(t)\right\}.
\label{eq:realizable_service}
\end{equation}
The term $c_m s_m$ is the signal-enabled capacity, whereas $\mu_m(t)$ limits service to traffic that is available upstream and can be received downstream. Under the active signal state, the stop-line outflow satisfies $q_m^{\mathrm{out}}(s_m^{\mathrm{out}}(t))=d_m(\bm s(t),t)$.

Traffic propagates to downstream movements according to turning proportions. Let $\mathcal U(m)$ denote the set of upstream movements feeding movement $m$, and let $r_{\ell m}(t)$ be the proportion of traffic discharged from $\ell\in\mathcal U(m)$ that subsequently enters $m$. The resulting inflow is
\begin{equation}
q_m^{\mathrm{in}}\!\left(s_m^{\mathrm{in}}(t)\right)
=
\sum_{\ell\in\mathcal U(m)}
r_{\ell m}(t)
d_\ell(\bm s(t),t),
\qquad
m\in\mathcal M_{\mathcal N}.
\label{eq:inflow_coupling}
\end{equation}
Equations~\eqref{eq:continuity_pde}--\eqref{eq:inflow_coupling} therefore map a signal state to realizable movement discharge and propagate that discharge through the network. The same turning structure is used later to construct ecological pressure. In implementation, the turning proportions are estimated from the intended movements reported by CVs \citep{tan2026cv}.

\subsection{Vehicle longitudinal dynamics and positive propulsion work}\label{sec:vehicle_power_work}

The vehicle-level ecological quantity is constructed from longitudinal motion so that a common state can be used across different powertrain types. The force model first determines tractive demand. Eco-MP then combines auxiliary power with the positive part of tractive power to define the positive propulsion power used in the movement state. Fuel consumption for PVs and electricity consumption for EVs are evaluated separately from the resulting trajectories.

For vehicle $k$ at time $t$, let $v_t^k$ and $a_t^k$ denote its speed and longitudinal acceleration. The parameters $M^k$, $A_f^k$, and $C_D^k$ denote vehicle mass, frontal area, and aerodynamic drag coefficient, respectively; $\mu_{\mathrm r}^k$ is the rolling-resistance coefficient, $\delta^k$ accounts for rotational inertia, and $\theta_t^k$ is the road-grade angle. The longitudinal force components are
\begin{align}
F_{\mathrm{grav},t}^k
&=M^k g\sin\theta_t^k,
\label{eq:force_gravity}\\
F_{\mathrm{roll},t}^k
&=\mu_{\mathrm r}^k M^k g\cos\theta_t^k,
\label{eq:force_rolling}\\
F_{\mathrm{air},t}^k
&=\frac{1}{2}\rho_a C_D^k A_f^k(v_t^k)^2,
\label{eq:force_aero}\\
F_{\mathrm{iner},t}^k
&=\delta^k M^k a_t^k,
\label{eq:force_inertia}
\end{align}
where $g$ is gravitational acceleration and $\rho_a$ is air density. These terms represent grade resistance, rolling resistance, aerodynamic drag, and acceleration-related inertial demand, respectively.

The total tractive-force demand and the corresponding mechanical power at the wheels are
\begin{align}
F_{\mathrm{trac},t}^k
&=
F_{\mathrm{grav},t}^k
+F_{\mathrm{roll},t}^k
+F_{\mathrm{air},t}^k
+F_{\mathrm{iner},t}^k,
\label{eq:force_trac}\\
P_{\mathrm{trac},t}^k
&=F_{\mathrm{trac},t}^k v_t^k.
\label{eq:mechanical_power}
\end{align}
Thus, wheel-level mechanical power is obtained by multiplying the total tractive force by vehicle speed.

Let $P_{\mathrm{aux},t}^k$ denote the auxiliary-power demand of vehicle $k$. Eco-MP defines the common positive propulsion power as
\begin{equation}
P_t^{k,+}
=
P_{\mathrm{aux},t}^k
+
\max\!\left\{0,P_{\mathrm{trac},t}^k\right\}.
\label{eq:positive_power}
\end{equation}
The corresponding cumulative positive propulsion work over an interval $[t_1,t_2]$ is
\begin{equation}
W^k(t_1,t_2)
=
\int_{t_1}^{t_2}P_\tau^{k,+}\,d\tau.
\label{eq:positive_work}
\end{equation}
This definition is common to PVs and EVs, allowing their contributions to be aggregated within the same movement state. Section~\ref{sssec:queue_memory_contribution} specifies when this accumulation begins and ends for each movement.

\section{Eco-MP Framework}\label{sec:framework}

Eco-MP coordinates signal service and vehicle motion through two coupled control layers in a mixed-powertrain environment. Let $T_i$ denote an MP signal-decision epoch. Once a green phase is selected, it is retained for the signal-control interval $\Delta T_s$; if the selected phase changes, a prescribed yellow interval is inserted before the next green. Let $t_i$ denote a vehicle-control epoch, with consecutive updates separated by $\Delta t$. At each $t_i$, a controlled CAV solves an $H$-step trajectory problem with prediction step $\Delta t$, implements only the first optimized acceleration command, and resolves the problem at $t_{i+1}=t_i+\Delta t$. The subscripts of $T_i$ and $t_i$ index the two update sequences independently, consistent with Figs.~\ref{fig:receding_signal_prediction} and \ref{fig:signal_prediction_cases}. The signal layer uses CV observations to construct ecological pressure and select phases, whereas the vehicle layer directly controls the CAV subset.

\begin{figure}[pos=t]
    \centering
    \includegraphics[width=\linewidth]{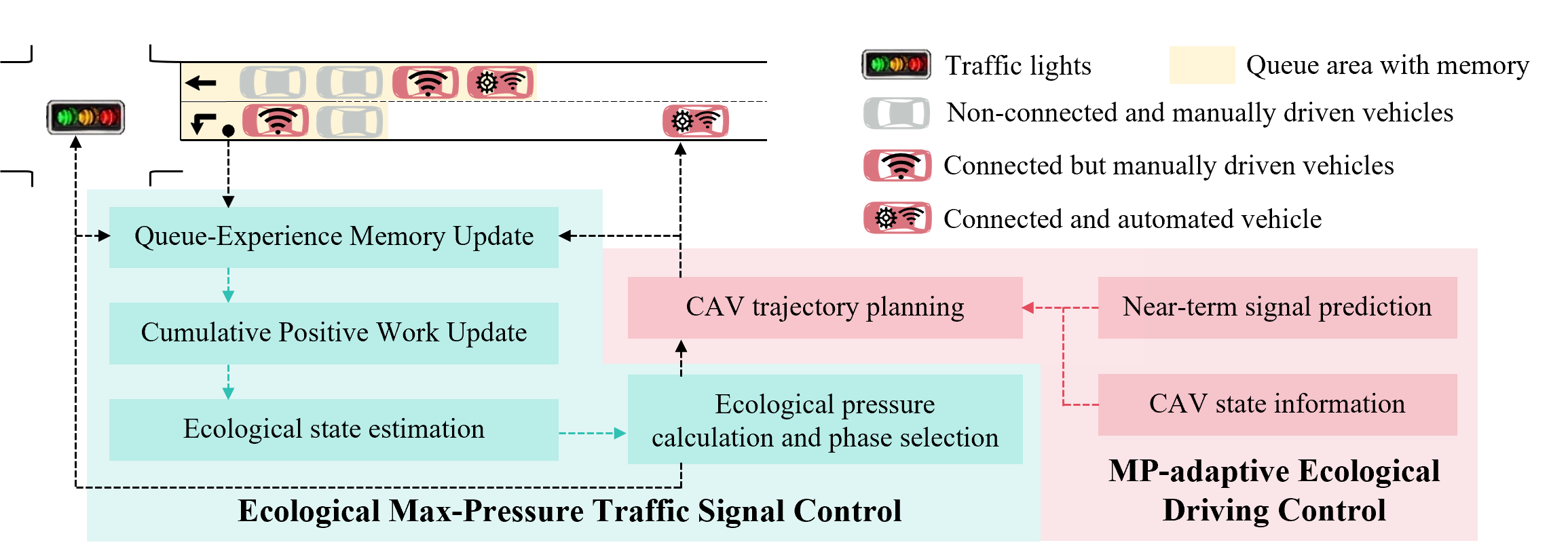}
    \caption{Closed-loop interaction between ecological MP signal control and MP-adaptive ecological driving.}
    \label{fig:eco_mp_framework}
\end{figure}

Figure~\ref{fig:eco_mp_framework} summarizes the closed-loop interaction. At each vehicle-control epoch, CV observations update queue experience, positive propulsion work, movement states, and ecological pressures. When an MP signal decision is due, the current ecological pressures determine the selected phase. Each controlled CAV then constructs a near-term signal description, solves its finite-horizon trajectory problem, and applies the first acceleration command. The resulting traffic evolution updates the states used by subsequent signal and vehicle-control decisions.

\subsection{Ecological MP signal control}\label{ssec:Eco-MP}

The signal layer retains the upstream--downstream comparison of classical MP control but replaces queue- or travel-time-based movement states with accumulated positive propulsion work. It records whether a vehicle has entered the detected queue region and retains that vehicle's contribution until discharge. The retained contributions are aggregated at the movement level, after which ecological pressure determines the signal phase.

\subsubsection{Queue-experience memory and vehicle contribution}\label{sssec:queue_memory_contribution}

\begin{figure}[pos=t]
    \centering
    \includegraphics[width=\linewidth]{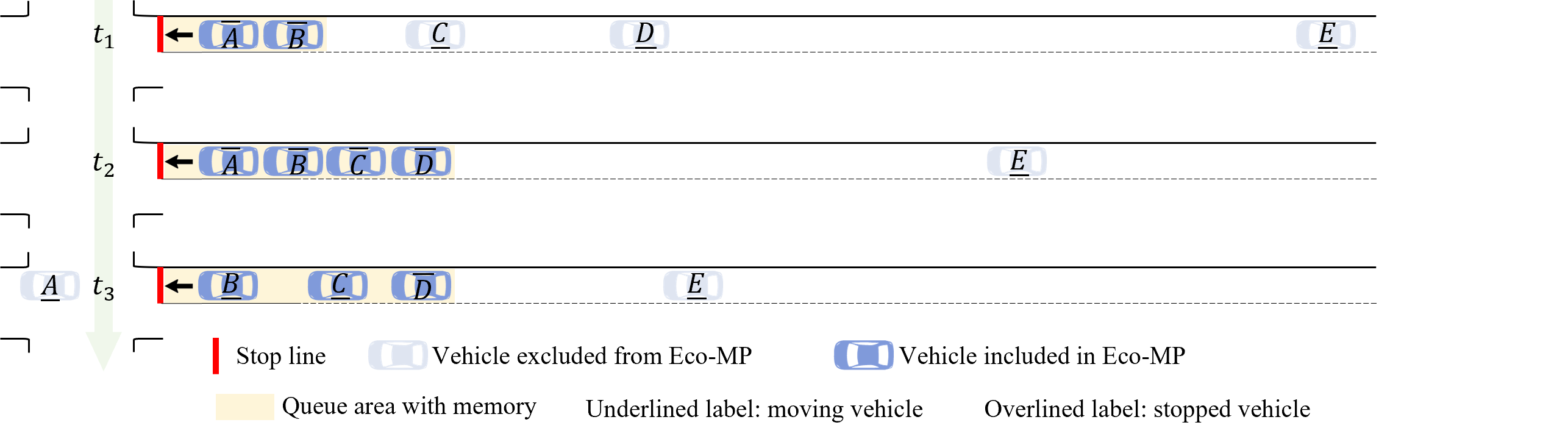}
    \caption{Queue-experience memory from first entry into the detected queue region to stop-line discharge. A vehicle remains active in the ecological state after it starts moving with a dissipating queue and is removed only after crossing the stop line.}
    \label{fig:queue_memory}
\end{figure}

Figure~\ref{fig:queue_memory} illustrates how queue experience is retained in the ecological state. For movement $m$, let $\mathcal K_m^{\mathrm{CV}}(t)$ denote the CVs assigned to $m$ that have not yet crossed its stop line; movement assignments follow their reported turning intentions. Let $v^{\mathrm q}$ denote the speed threshold used to identify low-speed vehicles. At time $t$, the farthest-upstream CV with speed below $v^{\mathrm q}$ defines the detected queue tail, and the segment from this queue tail to the stop line is treated as the current queue region. If no such CV is observed, the current queue region is empty.

Let $\beta_m^k(t)\in\{0,1\}$ indicate whether vehicle $k$ has entered the detected queue region during its current traversal of movement $m$ and has not yet crossed the stop line:
\begin{equation}
\beta_m^k(t)
=
\begin{cases}
1, & \text{if vehicle }k\text{ has entered the detected queue region and remains upstream of the stop line},\\
0, & \text{otherwise}.
\end{cases}
\label{eq:beta_update}
\end{equation}
Once activated, this memory remains active while the vehicle is upstream of the stop line, even if the queue dissipates and the vehicle accelerates above $v^{\mathrm q}$; it is cleared after stop-line discharge. The vehicles carrying an ecological contribution are therefore $\mathcal K_m^{\mathrm{eco}}(t)=\{k\in\mathcal K_m^{\mathrm{CV}}(t):\beta_m^k(t)=1\}$.

This design focuses the ecological state on propulsion work associated with signal-induced queueing and discharge. Propulsion work during uninterrupted progression, particularly at higher speeds, can be substantial but is less directly associated with queue formation and signal delay. Once a vehicle enters the queue region, however, its subsequent stopping, restarting, and acceleration toward the stop line remain part of the same queueing-and-discharge episode. Its contribution is therefore accumulated from first entry into the detected queue region until stop-line discharge.

Let $t_{m,q}^k$ denote the first time at which vehicle $k$ enters the detected queue region during its current traversal of movement $m$. Its movement-specific cumulative contribution is
\begin{equation}
W_m^k(t)
=
W^k(t_{m,q}^k,t)
=
\int_{t_{m,q}^k}^{t}P_\tau^{k,+}\,d\tau,
\qquad
k\in\mathcal K_m^{\mathrm{eco}}(t).
\label{eq:vehicle_positive_work_movement}
\end{equation}
The contribution continues to accumulate while the vehicle remains upstream of the stop line and is removed after the vehicle discharges through movement $m$.

\subsubsection{Ecological pressure and signal decision}\label{sssec:eco_pressure_decision}

The memory rule determines which vehicles remain associated with the current stop-and-go episode; their accumulated contributions are then aggregated into the movement state used for signal control. For movement $m=(i,o)$,
\begin{equation}
\xi_m(t)
=
\frac{1}{L_i}
\sum_{k\in\mathcal K_m^{\mathrm{eco}}(t)}
W_m^k(t).
\label{eq:movement_eco_state}
\end{equation}
The numerator is the cumulative positive propulsion work carried by CVs that have entered the detected queue region. Division by the incoming-link length $L_i$ normalizes the state across movements with different spatial extents. Thus, $\xi_m$ reflects both the number of vehicles affected by queueing and the positive propulsion work accumulated after queue entry.

Eco-MP compares the ecological state awaiting service with the state expected downstream. Let $\mathcal D(m)$ denote the set of downstream movements reachable from movement $m$. The turning proportion $\hat r_{mm'}(t)$ from $m$ to $m'$ is estimated from the intended downstream movements reported by CVs using the connected-vehicle turning-estimation procedure in \citet{tan2026cv}. These proportions weight downstream ecological states according to the expected destinations of discharged vehicles. The movement ecological pressure is
\begin{equation}
\pi_m(t)
=
\xi_m(t)
-
\sum_{m'\in\mathcal D(m)}
\hat r_{mm'}(t)\xi_{m'}(t).
\label{eq:eco_movement_pressure}
\end{equation}
The first term is the ecological burden awaiting service on movement $m$, while the second is the turning-weighted downstream burden expected after discharge. A large positive $\pi_m$ therefore indicates high upstream ecological burden relative to the expected downstream burden.

For a feasible phase $p\in\mathcal P_n$ at intersection $n$, let $s_m^p$ indicate whether movement $m\in\mathcal M_n$ is served by that phase. Its realizable service under phase $p$ is
$d_m^p(t)=\min\{c_m s_m^p,\mu_m(t)\}$,
consistent with Eq.~\eqref{eq:realizable_service}. The phase pressure and the corresponding Eco-MP decision are
\begin{align}
\Pi_p(t)
&=
\sum_{m\in\mathcal M_n}
d_m^p(t)\pi_m(t),
\label{eq:eco_phase_pressure}\\
p_n^\star(t)
&=
\arg\max_{p\in\mathcal P_n}\Pi_p(t).
\label{eq:eco_mp_rule}
\end{align}
The weight $d_m^p(t)$ accounts for the amount of movement $m$ that can actually discharge under phase $p$, so $\Pi_p$ measures the ecological pressure relief realizable by that phase. Equations~\eqref{eq:vehicle_positive_work_movement}--\eqref{eq:eco_mp_rule} therefore define a sequential signal-control mechanism: queue experience determines the active vehicle contributions, these contributions form $\xi_m$, turning-weighted downstream states convert $\xi_m$ into movement pressure $\pi_m$, and realizable discharge converts movement pressure into the phase objective $\Pi_p$. Eco-MP thus ranks phases by the ecological pressure that can be served under the current traffic state.

\subsection{MP-adaptive ecological driving}\label{ssec:Eco-driving}

The vehicle layer controls CAV longitudinal motion to reduce unnecessary approach and discharge maneuvers. Eco-MP determines future phases from evolving ecological pressures, and each CAV receives a rolling near-term description of this signal process before solving its finite-horizon trajectory problem. CAV trajectories are optimized independently, with interactions with preceding vehicles represented through the car-following constraint introduced in Section~\ref{sssec:finite_horizon_control}.

\subsubsection{Rolling near-term signal prediction}\label{sssec:rolling_signal_prediction}

At each vehicle-control epoch $t_i$, the vehicle layer constructs a near-term signal description for the approaching movement using the current signal indication, the predicted remaining time to the next effective green--red change, and the indication after that change. The prediction is initialized from the signal and ecological states observed at $t_i$. If the prediction horizon extends across a scheduled MP decision, the next phase is predicted by applying the Eco-MP phase-selection rule to the latest ecological state available at $t_i$, together with the prescribed phase-transition logic. Because the description is recomputed every $\Delta t$, this prediction is updated as new ecological-state information becomes available. The yellow interval is included in the predicted transition timing and is treated as red for trajectory planning.

For CAV $k$, let $x_i^k=x^k(t_i)$ denote its longitudinal position along the approach at the current control epoch, and let $x_{\mathrm{stop}}^k$ denote the stop-line coordinate in the same reference system. The current distance to the stop line is therefore $x_{\mathrm{stop}}^k-x_i^k$. Define the near-term signal tuple as $\mathcal I_i^k=(\chi_i^k,\tau_i^{\mathrm{sw},k},\chi_i^{\mathrm{next},k})$, where $\chi_i^k\in\{\mathrm{G},\mathrm{R}\}$ is the current effective indication, $\tau_i^{\mathrm{sw},k}$ is the predicted remaining time to the next effective change, and $\chi_i^{\mathrm{next},k}\in\{\mathrm{G},\mathrm{R}\}$ is the indication after that change. Vehicles approaching the same movement share the same signal evolution, whereas their positions, speeds, and leader conditions remain vehicle specific.

\begin{figure}[pos=t]
    \centering
    \includegraphics[width=\linewidth]{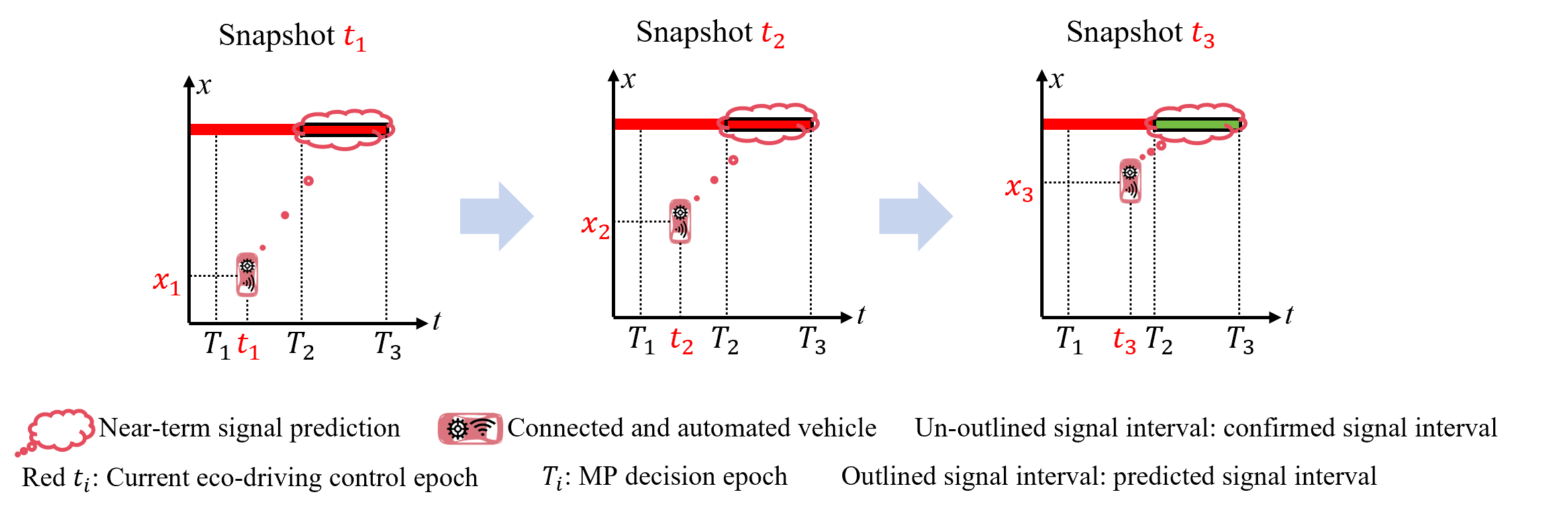}
    \caption{Receding-horizon signal prediction and vehicle control between two consecutive Eco-MP decision epochs.}
    \label{fig:receding_signal_prediction}
\end{figure}

The tuple is recomputed every $\Delta t$, allowing the vehicle trajectory to be replanned multiple times between two MP signal decisions. Figure~\ref{fig:receding_signal_prediction} uses $T_i$ for MP decision epochs, $t_i$ for vehicle-control epochs, and $x_i=x^k(t_i)$ for the vehicle position at successive updates.

\subsubsection{Finite-horizon optimal-control formulation}\label{sssec:finite_horizon_control}

\begin{figure}[pos=t]
    \centering
    \includegraphics[width=\linewidth]{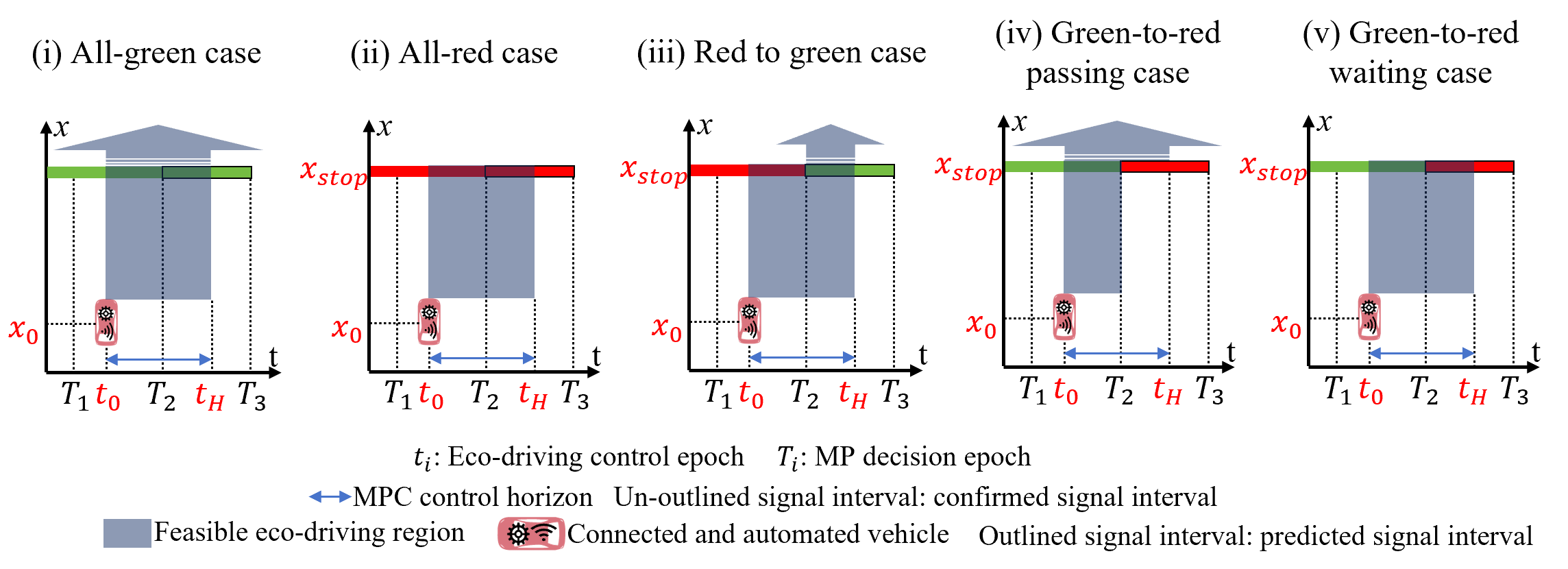}
    \caption{Signal-sequence cases used to construct the position constraints of the ecological-driving problem.}
    \label{fig:signal_prediction_cases}
\end{figure}

At vehicle-control epoch $t_i$, CAV $k$ solves an $H$-step quadratic program (QP) with time step $\Delta t$. Prediction index $\ell=0,\ldots,H$ corresponds to time $t_i+\ell\Delta t$. The predicted state is $\bm z_\ell^k=[x_\ell^k,v_\ell^k]^\top$, where $x_\ell^k$ is longitudinal position in the same approach coordinate used in Figs.~\ref{fig:receding_signal_prediction} and \ref{fig:signal_prediction_cases}, and $v_\ell^k$ is speed. The measured initial state is $\bm z_0^k=[x_0^k,v_0^k]^\top=[x_i^k,v^k(t_i)]^\top$. The control input is longitudinal acceleration, $u_\ell^k=a_\ell^k$, and $\bm u^k=[u_0^k,\ldots,u_{H-1}^k]^\top$ denotes the optimized control sequence.

The tuple $\mathcal I_i^k$ selects the signal case in Fig.~\ref{fig:signal_prediction_cases}, while the stop-line coordinate $x_{\mathrm{stop}}^k$ provides the corresponding spatial boundary. Together with the measured ego state and leader prediction, these quantities define the time-varying inputs to
\begin{equation}
\begin{aligned}
\min_{\bm u^k}\quad &J^k(\bm u^k)\\
\mathrm{s.t.}\quad
&\text{vehicle dynamics in Eq.~\eqref{eq:eco_driving_dynamics}},\\
&\text{acceleration and speed bounds in Eqs.~\eqref{eq:eco_driving_acc_constraint}--\eqref{eq:eco_driving_speed_constraint}},\\
&\text{car-following bound in Eq.~\eqref{eq:eco_driving_safety_constraint}},\\
&\text{signal constraint selected from Eqs.~\eqref{eq:sig_red_red}--\eqref{eq:sig_green_red_wait}}\\
&\text{according to }\mathcal I_i^k.
\end{aligned}
\label{eq:eco_driving_qp}
\end{equation}
The objective balances progress toward the desired speed against acceleration effort; the vehicle dynamics propagate each candidate trajectory; the acceleration, speed, and car-following constraints define admissible longitudinal motion; and the signal constraint determines whether the vehicle may cross the stop line within the predicted green interval or must remain upstream. These components are specified below.

\paragraph{Vehicle dynamics.}
Under constant acceleration within one prediction interval, the state evolves as
\begin{equation}
\bm z_{\ell+1}^k
=
A\bm z_\ell^k+Bu_\ell^k,
\qquad
A=
\begin{bmatrix}
1&\Delta t\\
0&1
\end{bmatrix},
\quad
B=
\begin{bmatrix}
\frac{1}{2}\Delta t^2\\
\Delta t
\end{bmatrix}.
\label{eq:eco_driving_dynamics}
\end{equation}
The first row updates longitudinal position from the current speed and acceleration, while the second updates speed. The initial position $x_0^k=x_i^k$ is the current vehicle position denoted by $x_0$ in Fig.~\ref{fig:signal_prediction_cases}.

\paragraph{Objective.}
The objective penalizes deviations from the desired speed and acceleration effort:
\begin{equation}
\begin{aligned}
J^k(\bm u^k)
={}&
\frac{1}{2}
\sum_{\ell=0}^{H-1}
\left[
q^{\mathrm{spd}}(v_\ell^k-v^{\mathrm{des}})^2
+
r^a(u_\ell^k)^2
\right]
+
\frac{1}{2}
q^{\mathrm{spd}}(v_H^k-v^{\mathrm{des}})^2,
\end{aligned}
\label{eq:eco_driving_objective}
\end{equation}
where $v^{\mathrm{des}}$ is the prescribed desired speed, $q^{\mathrm{spd}}\geq0$ weights speed-tracking error, and $r^a>0$ weights acceleration effort. The position trajectory follows from the vehicle dynamics together with the car-following and signal constraints. The acceleration penalty discourages abrupt speed corrections, while the signal constraint can move deceleration earlier when a stop is predicted. The same trajectory objective is used for PVs and EVs; fuel and electricity consumption are evaluated separately for the resulting trajectories.

\paragraph{Physical constraints.}
The acceleration and speed are bounded by
\begin{align}
a_{\min}
&\leq
u_\ell^k
\leq
a_{\max},
&& \ell=0,\ldots,H-1,
\label{eq:eco_driving_acc_constraint}\\
v_{\min}
&\leq
v_\ell^k
\leq
v_{\max},
&& \ell=0,\ldots,H.
\label{eq:eco_driving_speed_constraint}
\end{align}
Here, $a_{\min}$ and $a_{\max}$ are the minimum and maximum admissible accelerations, while $v_{\min}$ and $v_{\max}$ are the minimum and maximum speeds. Longitudinal position evolves according to Eq.~\eqref{eq:eco_driving_dynamics}; the signal constraints below determine whether the vehicle may cross the stop line.

\paragraph{Car-following constraint.}
To account for predicted leader motion, the optimized acceleration is upper-bounded by an intelligent driver model (IDM) term:
\begin{align}
u_\ell^k
&\leq
\bar a_{\mathrm{IDM},\ell}^k,
&& \ell=0,\ldots,H-1,
\label{eq:eco_driving_safety_constraint}\\
\bar a_{\mathrm{IDM},\ell}^k
&=
a_{\max}
\left[
1-
\left(\frac{\bar v_\ell^k}{v_{\mathrm{free}}^{\mathrm{IDM}}}\right)^{\delta_{\mathrm{IDM}}}
-
\left(\frac{d_\ell^{*,k}}{\bar d_\ell^{\mathrm{gap},k}}\right)^2
\right],
\label{eq:idm_acc}\\
d_\ell^{*,k}
&=
d_0+\bar v_\ell^k T_h
+
\frac{\bar v_\ell^k(\bar v_\ell^k-\bar v_\ell^{\mathrm{lead},k})}
{2\sqrt{a_{\max}b}}.
\label{eq:idm_desired_gap}
\end{align}
Here, $v_{\mathrm{free}}^{\mathrm{IDM}}$ and $\delta_{\mathrm{IDM}}$ are the IDM free-flow speed and exponent; $\bar v_\ell^k$, $\bar v_\ell^{\mathrm{lead},k}$, and $\bar d_\ell^{\mathrm{gap},k}$ are the ego speed, leader speed, and inter-vehicle gap used to evaluate the IDM bound; and $d_0$, $T_h$, and $b$ are the standstill gap, desired headway, and comfortable-deceleration magnitude. These quantities are updated from the latest ego and leader predictions before each QP solve. The resulting sequence $\bar a_{\mathrm{IDM},\ell}^k$ is held fixed during that solve, so Eq.~\eqref{eq:eco_driving_safety_constraint} remains affine in the optimization variables. If no relevant leader is present, the bound is inactive.

\paragraph{Signal constraints.}

Figure~\ref{fig:signal_prediction_cases} uses $x_0$ for the vehicle position at the current vehicle-control epoch and $x_{\mathrm{stop}}$ for the stop-line coordinate. In the vehicle-specific formulation, these correspond to $x_0^k=x_i^k$ and $x_{\mathrm{stop}}^k$, defined in the same longitudinal coordinate system. Let $\tau_\ell=\ell\Delta t$ denote elapsed prediction time. If the signal remains green throughout the relevant prediction interval, no additional stop-line constraint is imposed. The remaining cases in Fig.~\ref{fig:signal_prediction_cases} are represented by
\begin{align}
x_\ell^k
&\leq
x_{\mathrm{stop}}^k,
&& \ell=0,\ldots,H,
&& \text{red remains red},
\label{eq:sig_red_red}\\
x_\ell^k
&\leq
x_{\mathrm{stop}}^k,
&& \tau_\ell\leq \tau_i^{\mathrm{sw},k},
&& \text{red to green},
\label{eq:sig_red_green}\\
x_{\ell_g}^k
&\geq
x_{\mathrm{stop}}^k,
&&
&& \text{green to red: pass},
\label{eq:sig_green_red_pass}\\
x_\ell^k
&\leq
x_{\mathrm{stop}}^k,
&& \ell=0,\ldots,H,
&& \text{green to red: wait},
\label{eq:sig_green_red_wait}
\end{align}
where $\ell_g=\lfloor\tau_i^{\mathrm{sw},k}/\Delta t\rfloor$ is the latest prediction index not later than the end of the current green. In the red-to-green case, the stop-line bound is released after the predicted green begins. In the green-to-red case, the pass formulation requires the vehicle to reach the stop line by the latest prediction point within the current green; if this problem is infeasible, the wait formulation keeps the vehicle at or upstream of the stop line over the horizon.

For a fixed signal case and a fixed IDM bound during the current solve, Eq.~\eqref{eq:eco_driving_qp} has a convex quadratic objective and affine constraints.

\subsubsection{Receding-horizon and closed-loop implementation}\label{sssec:receding_horizon}\label{ssec:closed_loop_implementation}

After solving Eq.~\eqref{eq:eco_driving_qp} at $t_i$, only the first optimized command, $a^k(t_i)=u_0^{k,\star}$, is implemented. At $t_{i+1}=t_i+\Delta t$, the vehicle state, leader prediction, and signal tuple are updated, and a new $H$-step QP is solved. This receding-horizon implementation continuously adapts the planned trajectory as traffic and signal conditions evolve.

The vehicle update operates together with the Eco-MP signal update to form the complete closed loop. The selected green phase is retained for $\Delta T_s$, subject to the prescribed yellow transition when the phase changes, while vehicle trajectories are replanned every $\Delta t$. Algorithm~\ref{alg:eco_mp_closed_loop} summarizes the execution sequence.

\begin{algorithm}
\caption{Closed-loop implementation of Eco-MP}\label{alg:eco_mp_closed_loop}
\begin{algorithmic}[1]
\Require Signal-control interval $\Delta T_s$, vehicle-control and prediction time step $\Delta t$, prediction horizon $H$, feasible phase sets $\{\mathcal P_n\}$
\State Initialize signal states, queue memories $\beta_m^k$, and movement contributions $W_m^k$
\For{each vehicle-control epoch $t_i$}
    \State Collect CV positions, speeds, accelerations, and turning intentions
    \State Identify the queue region of each movement and update $\beta_m^k(t_i)$ using Eq.~\eqref{eq:beta_update}
    \State Compute $P_{t_i}^{k,+}$ using Eq.~\eqref{eq:positive_power} and update $W_m^k(t_i)$ using Eq.~\eqref{eq:vehicle_positive_work_movement}
    \State Compute $\xi_m(t_i)$ using Eq.~\eqref{eq:movement_eco_state}, update $\hat r_{mm'}(t_i)$, and compute $\pi_m(t_i)$ using Eq.~\eqref{eq:eco_movement_pressure}
    \If{a new MP signal decision is due}
        \ForAll{real intersections $n$}
            \State Evaluate $d_m^p(t_i)$ and $\Pi_p(t_i)$ for each $p\in\mathcal P_n$
            \State Select $p_n^\star(t_i)$ using Eq.~\eqref{eq:eco_mp_rule} and apply the prescribed phase-transition rule
        \EndFor
    \EndIf
    \ForAll{controlled CAVs $k$}
        \State Construct $\mathcal I_i^k$ from the current signal and ecological states
        \State Update the IDM bounds using Eqs.~\eqref{eq:idm_acc}--\eqref{eq:idm_desired_gap}
        \State Identify the signal case in Fig.~\ref{fig:signal_prediction_cases}
        \If{the signal remains green}
            \State Solve Eq.~\eqref{eq:eco_driving_qp} without an additional stop-line constraint
        \ElsIf{the signal remains red}
            \State Solve Eq.~\eqref{eq:eco_driving_qp} with Eq.~\eqref{eq:sig_red_red}
        \ElsIf{the signal changes from red to green}
            \State Solve Eq.~\eqref{eq:eco_driving_qp} with Eq.~\eqref{eq:sig_red_green}
        \ElsIf{the signal changes from green to red}
            \State Solve Eq.~\eqref{eq:eco_driving_qp} with Eq.~\eqref{eq:sig_green_red_pass}
            \If{the QP is infeasible}
                \State Solve Eq.~\eqref{eq:eco_driving_qp} with Eq.~\eqref{eq:sig_green_red_wait}
            \EndIf
        \EndIf
        \State Apply only the first command $a^k(t_i)=u_0^{k,\star}$
    \EndFor
    \State Advance the traffic state to $t_{i+1}=t_i+\Delta t$
\EndFor
\end{algorithmic}
\end{algorithm}

Algorithm~\ref{alg:eco_mp_closed_loop} completes the interaction between the two layers: Eco-MP updates signal decisions from the current ecological state, while CAV trajectories respond to the resulting near-term signal process. The resulting traffic evolution then updates the states used at subsequent control epochs.

\section{Stability Analysis}\label{sec:stability}

This section establishes network-stability properties for Eco-MP and the coupled signal--vehicle framework. The analysis follows the finite-storage setting used for max-pressure control with finite physical links \citep{li2019position,tan2026cv}. Real-link occupancies are physically bounded, fictitious source queues represent unmet boundary demand, and the realizable movement service used by Eco-MP appears explicitly in the Lyapunov drift.

For the stability analysis, turning proportions are treated as fixed network parameters. We analyze the idealized Eco-MP phase-selection rule evaluated at the current traffic state. The finite green-holding and yellow-transition intervals used in the microscopic implementation are not included in this stability result. The feasible network signal set factorizes across intersections, so a network signal state consists of one feasible local phase at each real intersection, while the fictitious-source service indicators remain fixed. The ecological states in the analysis are their full-state counterparts; the relation to sampled-CV observations is stated after the two stability results.

\subsection{Stability setting and admissible demand region}\label{ssec:stability_preliminaries}

Throughout this section, $a$ indexes an arbitrary movement, $m$ a real movement in $\mathcal M_{\mathcal N}$, and $f$ a fictitious source movement in $\mathcal M_{\mathcal F}$. For a real movement $m=(i,o)$, let $z_m(t)=\int_0^{L_i}\rho_m(x,t)\,dx$ denote its vehicle count; for a fictitious source movement, let $z_f(t)=\rho_f(t)$ denote the corresponding point-queue length. Finite physical storage implies
\begin{equation}
0\leq z_m(t)\leq\bar z_m<\infty,
\qquad
m\in\mathcal M_{\mathcal N}.
\label{eq:finite_real_storage}
\end{equation}

\begin{definition}[Strong network stability]\label{def:strong_stability}
The traffic network is strongly stable if
\[
\limsup_{T\rightarrow\infty}
\frac{1}{T}
\int_0^T
\mathbb E\!\left[
\sum_{a\in\mathcal M_{\mathcal N}\cup\mathcal M_{\mathcal F}}z_a(t)
\right]dt
<\infty.
\]
\end{definition}

Because Eq.~\eqref{eq:finite_real_storage} uniformly bounds every real-movement count, strong network stability follows once the fictitious source queues have finite long-run average accumulation.

Let $\bm r=[r_{ab}]$ be the fixed routing matrix, where $r_{ab}$ is the fraction of discharge from movement $a$ that subsequently enters movement $b$. The matrix is nonnegative and row-substochastic, with missing row mass representing traffic that leaves the modeled network. The network is assumed to be open, so its spectral radius satisfies $\varrho(\bm r)<1$.

For a feasible network signal state $\bm s\in\mathbb S$, let $d_a(\bm s,t)$ denote the realizable service of movement $a$ at time $t$. For a real movement, this is given by Eq.~\eqref{eq:realizable_service}; for a fictitious source movement, discharge represents admission into the physical network. The corresponding net accumulation-rate vector is
\begin{equation}
\bm h(\bm s,t)
=
\bm\lambda(t)+\bm r^\top\bm d(\bm s,t)-\bm d(\bm s,t),
\label{eq:net_accumulation}
\end{equation}
where $\bm d(\bm s,t)$ collects the movement services and $\bm\lambda(t)$ has nonzero entries only for fictitious source movements.

Following the network-region constructions in \citet{li2019position} and \citet{tan2026cv}, define the set of mean arrival rates supportable by a network-wide signal scheme $\bm s$ as
\begin{equation}
\Omega(\bm s)
=
\left\{
\bar{\bm\lambda}\geq\bm0:
\limsup_{T\rightarrow\infty}
\frac{1}{T}
\int_0^T
\mathbb E[\bm h(\bm s,t)]\,dt
\preceq\bm0
\right\}.
\label{eq:admissible_flow_set}
\end{equation}
The admissible demand region is then defined as
\begin{definition}[Admissible demand region]\label{def:admissible_demand_region}
\begin{equation}
\Lambda
=
\operatorname{conv}
\left(
\bigcup_{\bm s\in\mathbb S}\Omega(\bm s)
\right).
\label{eq:admissible_demand_region}
\end{equation}
\end{definition}
The convex-hull construction represents combinations of feasible network signal schemes in the standard MP region formulation. A network signal policy is a feasible curve $\bm s(\cdot)$ taking values in $\mathbb S$. The stability result below considers demands in $\Lambda^{\mathrm{int}}$ that additionally satisfy the strict supportability condition derived from the Lyapunov drift.

Let $\mathcal H_{t^-}$ denote the traffic and control information available immediately before time $t$, and assume $\mathbb E[\bm\lambda(t)\mid\mathcal H_{t^-}]=\bar{\bm\lambda}$. Vehicle and environmental parameters, vehicle speed and acceleration, and auxiliary power are bounded. Equations~\eqref{eq:force_gravity}--\eqref{eq:positive_power} therefore imply a finite constant $\bar P_m$ such that $0\leq P_t^{k,+}\leq\bar P_m$ for every vehicle represented in the ecological state of movement $m$.

\subsection{Ecological Lyapunov function and Eco-MP stability}\label{ssec:eco_mp_stability}

For the full-state analysis, let $\mathcal K_m(t)$ denote all vehicles on movement $m$. The queue region and $\beta_m^k$ follow the same rule as in Section~\ref{sssec:queue_memory_contribution}, now using complete vehicle observations. The normalized contribution carried by vehicle $k$ on movement $m=(i,o)$ is
\[
e_m^k(t)=
\begin{cases}
W_m^k(t)/L_i, & \beta_m^k(t)=1,\\
0, & \beta_m^k(t)=0,
\end{cases}
\qquad
\xi_m(t)=\sum_{k\in\mathcal K_m(t)}e_m^k(t),
\]
where $\xi_m$ is the full-state counterpart of Eq.~\eqref{eq:movement_eco_state}. A vehicle contribution is initialized at zero when queue memory is activated and is removed at stop-line discharge.

The positive-propulsion-work accumulation rate and its finite-storage bound are
\begin{equation}
\alpha_m(t)
=
\frac{1}{L_i}
\sum_{\substack{k\in\mathcal K_m(t)\\ \beta_m^k(t)=1}}
P_t^{k,+},
\qquad
0\leq\alpha_m(t)
\leq
\frac{\bar z_m\bar P_m}{L_i}
\triangleq
\bar\alpha_m
<\infty.
\label{eq:eco_accumulation_bound}
\end{equation}
If $g_m(t)\geq0$ denotes the rate at which accumulated contribution leaves with discharging vehicles, then $\dot\xi_m(t)=\alpha_m(t)-g_m(t)$.

The source-queue state $z_f$ and the ecological state $\xi_m$ have different physical units. Introduce a fixed analytical scaling factor $\kappa_E>0$ and define
\begin{equation}
w_a(t)
=
\begin{cases}
z_a(t), & a\in\mathcal M_{\mathcal F},\\
\kappa_E\xi_a(t), & a\in\mathcal M_{\mathcal N},
\end{cases}
\qquad
\varpi_a(t)
=
w_a(t)-\sum_b r_{ab}w_b(t).
\label{eq:analytical_pressure}
\end{equation}
For every real movement, $\varpi_m$ equals $\kappa_E$ times the full-state counterpart of the ecological pressure in Eq.~\eqref{eq:eco_movement_pressure}. Because $\kappa_E$ is positive and common to all real movements, it does not change the phase maximizer.

Consider the Lyapunov function
\begin{equation}
V(t)
=
\frac{1}{2}
\sum_{f\in\mathcal M_{\mathcal F}}
z_f^2(t)
+
\kappa_E
\sum_{m\in\mathcal M_{\mathcal N}}
z_m(t)\xi_m(t).
\label{eq:eco_lyapunov}
\end{equation}
The quadratic source term captures the unbounded boundary queues, while the matched term $\kappa_E z_m\xi_m$ couples each bounded real-movement occupancy with its accumulated ecological burden.

\begin{lemma}[Ecological Lyapunov drift]\label{lem:eco_drift}
Let $K_E=\kappa_E\sum_{m\in\mathcal M_{\mathcal N}}\bar z_m^2\bar P_m/L_i<\infty$. For every feasible signal state $\bm s$,
\begin{align}
\mathbb E\!\left[
\left.
\dot V(t)
\right|
\mathcal H_{t^-},\bm s
\right]
&\leq
K_E
+
\bm w^\top(t)\bar{\bm\lambda}
-
\Psi(\bm s,t),
\label{eq:eco_drift}\\
\Psi(\bm s,t)
&=
\sum_a
\varpi_a(t)d_a(\bm s,t).
\label{eq:network_eco_objective}
\end{align}
\end{lemma}

\begin{proof}[Main steps]
For a real movement, traffic conservation gives $\dot z_m=q_m^{\mathrm{in}}-d_m(\bm s,t)$, while $\dot\xi_m=\alpha_m-g_m$. Hence
\begin{align*}
\frac{d}{dt}
\left[
\kappa_E z_m\xi_m
\right]
&=
\kappa_E
\left(
\xi_m\dot z_m
+
z_m\dot\xi_m
\right)\\
&\leq
\kappa_E\xi_m
\left[
q_m^{\mathrm{in}}
-
d_m(\bm s,t)
\right]
+
\kappa_E z_m\alpha_m,
\end{align*}
because $-\kappa_E z_mg_m\leq0$. Equation~\eqref{eq:eco_accumulation_bound} bounds the last term by $\kappa_E\bar z_m^2\bar P_m/L_i$.

For a fictitious source, $d(z_f^2/2)/dt=z_f(\lambda_f-d_f)$. Using $q_m^{\mathrm{in}}=\sum_a r_{am}d_a(\bm s,t)$ and summing over all movements yields Eq.~\eqref{eq:eco_drift}. Queue-memory activation begins with zero accumulated work, and removing a vehicle contribution at discharge cannot increase the matched Lyapunov term. Appendix~\ref{app:proof_eco_drift} provides the complete derivation.
\end{proof}

Lemma~\ref{lem:eco_drift} links the ecological state directly to the signal decision. For a real intersection $n$, the phase-dependent part of the network objective is
\[
\Psi_n(p,t)
=
\sum_{m\in\mathcal M_n}
\varpi_m(t)d_m^p(t)
=
\kappa_E
\sum_{m\in\mathcal M_n}
\pi_m(t)d_m^p(t).
\]
All other terms in $\Psi$ are independent of the candidate phase at intersection $n$. Because the feasible network signal set factorizes across intersections, the collection of local Eco-MP decisions maximizes $\Psi(\bm s,t)$ over $\mathbb S$.

\begin{definition}[Strict supportability]\label{def:strict_supportability}
A mean demand $\bar{\bm\lambda}\in\Lambda^{\mathrm{int}}$ is strictly supportable if there exists a feasible reference signal policy $\bm s^{\circ}(\cdot)$ and constants $K^{\circ}<\infty$ and $\epsilon>0$ such that, for all $t$ and all admissible network states,
\begin{equation}
K_E
+
\bm w^\top(t)\bar{\bm\lambda}
-
\Psi(\bm s^{\circ}(t),t)
\leq
K^{\circ}
-
\epsilon
\sum_{f\in\mathcal M_{\mathcal F}}z_f(t).
\label{eq:reference_drift_margin}
\end{equation}
\end{definition}

\begin{theorem}[Stability of Eco-MP]\label{thm:eco_mp_stability}
Under the assumptions in Section~\ref{ssec:stability_preliminaries} and $\mathbb E[V(0)]<\infty$, Eco-MP strongly stabilizes the traffic network for every mean demand $\bar{\bm\lambda}\in\Lambda^{\mathrm{int}}$ that satisfies the strict supportability condition in Definition~\ref{def:strict_supportability}.
\end{theorem}

\begin{proof}[Main steps]
By Definition~\ref{def:strict_supportability}, there exists a feasible reference policy $\bm s^{\circ}(t)$ and constants $K^{\circ}<\infty$ and $\epsilon>0$ satisfying Eq.~\eqref{eq:reference_drift_margin}. Because Eco-MP maximizes $\Psi$,
\[
\Psi(\bm s^\star(t),t)
\geq
\Psi(\bm s^{\circ}(t),t).
\]
Combining this comparison with Lemma~\ref{lem:eco_drift} gives
\begin{equation}
\mathbb E\!\left[
\left.
\dot V(t)
\right|
\mathcal H_{t^-}
\right]
\leq
K^{\circ}
-
\epsilon
\sum_{f\in\mathcal M_{\mathcal F}}z_f(t).
\label{eq:eco_negative_drift}
\end{equation}
Integrating Eq.~\eqref{eq:eco_negative_drift} over $[0,T]$ yields
\[
\frac{1}{T}
\int_0^T
\mathbb E\!\left[
\sum_f z_f(t)
\right]dt
\leq
\frac{K^{\circ}}{\epsilon}
+
\frac{\mathbb E[V(0)]}{\epsilon T}.
\]
The fictitious source queues therefore have finite long-run average accumulation, and Eq.~\eqref{eq:finite_real_storage} bounds all real-movement counts. Strong network stability follows from Definition~\ref{def:strong_stability}. Appendix~\ref{app:proof_eco_stability} gives the full calculation.
\end{proof}

Theorem~\ref{thm:eco_mp_stability} shows that the ecological movement state preserves the MP stability mechanism: Eco-MP maximizes the same realizable pressure--service term that enters the network drift with a negative sign.

\subsection{Stability with MP-adaptive ecological driving}\label{ssec:eco_driving_stability}

When CAVs apply the controller in Section~\ref{ssec:Eco-driving}, their trajectories modify both the ecological state and the realizable departure process. Let $\tilde\xi_m(t)$ denote the resulting full-state ecological state of movement $m$, and let $\tilde d_a(\bm s,t)$ denote the corresponding realizable movement service. Define $\tilde\Omega(\bm s)$ by Eq.~\eqref{eq:admissible_flow_set} with $d_a$ replaced by $\tilde d_a$, and define the coupled admissible demand region as $\tilde\Lambda=\operatorname{conv}(\bigcup_{\bm s\in\mathbb S}\tilde\Omega(\bm s))$.

Define the coupled analytical state and pressure by
\[
\tilde w_a(t)=
\begin{cases}
z_a(t), & a\in\mathcal M_{\mathcal F},\\
\kappa_E\tilde\xi_a(t), & a\in\mathcal M_{\mathcal N},
\end{cases}
\qquad
\tilde\varpi_a(t)=\tilde w_a(t)-\sum_b r_{ab}\tilde w_b(t).
\]
The coupled pressure--service objective and Lyapunov function are
\begin{align}
\tilde\Psi(\bm s,t)
&=
\sum_a
\tilde\varpi_a(t)\tilde d_a(\bm s,t),
\label{eq:coupled_network_objective}\\
\tilde V(t)
&=
\frac{1}{2}\sum_f z_f^2(t)
+
\kappa_E\sum_m z_m(t)\tilde\xi_m(t).
\label{eq:coupled_lyapunov}
\end{align}

\begin{definition}[Strict supportability of the coupled process]\label{def:coupled_strict_supportability}
A mean demand $\bar{\bm\lambda}\in\tilde\Lambda^{\mathrm{int}}$ is strictly supportable for the coupled process if there exists a feasible reference policy $\tilde{\bm s}^{\circ}(\cdot)$ and constants $\tilde K^{\circ}<\infty$ and $\tilde\epsilon>0$ such that, for all $t$ and all admissible coupled-system states,
\begin{equation}
\tilde K_E
+
\tilde{\bm w}^\top(t)\bar{\bm\lambda}
-
\tilde\Psi(\tilde{\bm s}^{\circ}(t),t)
\leq
\tilde K^{\circ}
-
\tilde\epsilon
\sum_{f\in\mathcal M_{\mathcal F}}z_f(t),
\label{eq:coupled_reference_drift_margin}
\end{equation}
where $\tilde K_E<\infty$ is the coupled drift constant defined in the proof below.
\end{definition}

\begin{theorem}[Stability of the coupled framework]\label{thm:coupled_stability}
Suppose the assumptions in Section~\ref{ssec:stability_preliminaries} hold for the coupled traffic process, vehicle acceleration and speed satisfy Eqs.~\eqref{eq:eco_driving_acc_constraint}--\eqref{eq:eco_driving_speed_constraint}, and $\mathbb E[\tilde V(0)]<\infty$. Then the coupled Eco-MP and MP-adaptive ecological-driving system strongly stabilizes every mean demand $\bar{\bm\lambda}\in\tilde\Lambda^{\mathrm{int}}$ that satisfies the strict supportability condition in Definition~\ref{def:coupled_strict_supportability}.
\end{theorem}

\begin{proof}[Main steps]
Bounded vehicle speed and acceleration give a finite positive-power bound and hence a finite drift constant $\tilde K_E$. Repeating Lemma~\ref{lem:eco_drift} for the coupled traffic process gives
\[
\mathbb E[\dot{\tilde V}(t)\mid\mathcal H_{t^-},\bm s]
\leq
\tilde K_E
+
\tilde{\bm w}^\top(t)\bar{\bm\lambda}
-
\tilde\Psi(\bm s,t).
\]
By Definition~\ref{def:coupled_strict_supportability}, a feasible reference policy provides a strict drift margin. Because the coupled Eco-MP rule maximizes $\tilde\Psi$, comparison with that policy yields
\[
\mathbb E[\dot{\tilde V}(t)\mid\mathcal H_{t^-}]
\leq
\tilde K^{\circ}
-
\tilde\epsilon
\sum_f z_f(t).
\]
The integration argument of Theorem~\ref{thm:eco_mp_stability} then establishes strong stability. Appendix~\ref{app:proof_coupled_stability} gives the complete calculation.
\end{proof}

Theorem~\ref{thm:coupled_stability} extends the same pressure--service argument to the trajectory-modified traffic process. Ecological driving changes both the ecological state and realizable movement service, while Eco-MP continues to maximize the corresponding network pressure--service objective.

\begin{corollary}[Admissible-demand-region non-contraction under enlarged admissible-flow sets]\label{cor:service_region_noncontraction}
Let $\Omega(\bm s)$ and $\tilde\Omega(\bm s)$ denote the admissible-flow sets without and with MP-adaptive ecological driving, respectively. If
\[
\Omega(\bm s)\subseteq\tilde\Omega(\bm s),
\qquad
\forall\bm s\in\mathbb S,
\]
then $\Lambda\subseteq\tilde\Lambda$. The inclusion is strict whenever the coupled admissible-flow sets support at least one demand outside $\Lambda$.
\end{corollary}

\begin{proof}[Main step]
The result follows directly from the admissible-demand-region definitions:
\[
\operatorname{conv}
\left(
\bigcup_{\bm s\in\mathbb S}\Omega(\bm s)
\right)
\subseteq
\operatorname{conv}
\left(
\bigcup_{\bm s\in\mathbb S}\tilde\Omega(\bm s)
\right).
\]
Appendix~\ref{app:proof_service_region_noncontraction} gives the set argument explicitly.
\end{proof}

Corollary~\ref{cor:service_region_noncontraction} states the network-level implication when trajectory-level service improvements enlarge the admissible-flow sets. For a movement with effective discharge headway $h_a^{\mathrm{eff}}$, a shorter headway allows more vehicles to discharge during an available green. If such service changes enlarge the admissible-flow sets, the coupled admissible demand region cannot contract. The lower stop-line headway observed under Eco-MP+ED in Fig.~\ref{fig:controller_mobility} is consistent with this mechanism.

\begin{remark}[Observation scope]\label{rem:observation_scope}
Theorems~\ref{thm:eco_mp_stability} and \ref{thm:coupled_stability} use full-state ecological quantities and fixed routing proportions. A sampled-CV implementation inherits the corresponding phase-selection guarantee when its ecological-state and turning estimates produce the same phase decisions as the full-state controller.
\end{remark}

\section{Evaluation}\label{sec:evaluation}

The evaluation addresses three questions: whether ecological pressure improves ecological and traffic performance relative to conventional MP states, what additional benefit is provided by MP-adaptive ecological driving, and how the integrated framework responds to changes in connectivity and fleet powertrain composition. The controller comparison provides the main performance evidence, representative trajectories illustrate the vehicle-level mechanism, and two sensitivity experiments characterize the tested deployment conditions.

\subsection{Experimental settings}\label{ssec:eval_setup}

\begin{figure}[pos=t]
    \centering
    \includegraphics[width=\linewidth]{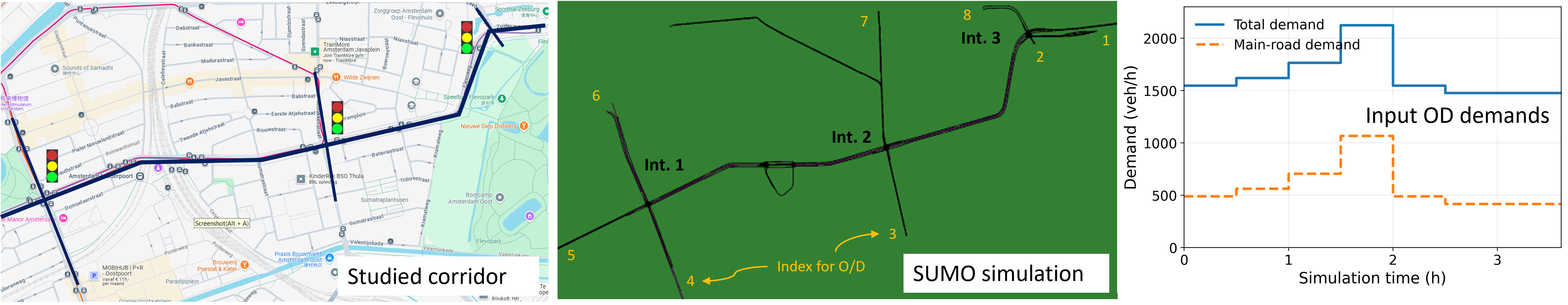}
    \caption{Amsterdam arterial corridor used in the simulation experiments.}
    \label{fig:test_network}
\end{figure}

A microscopic simulation environment was developed in Simulation of Urban MObility (SUMO) and interfaced with Python through the Traffic Control Interface (TraCI), allowing the signal and vehicle controllers to be updated online \citep{lopez2018microscopic}. The study network, extracted from OpenStreetMap, represents an urban arterial corridor in Amsterdam, the Netherlands. As shown in Fig.~\ref{fig:test_network}, the network has a total lane length of approximately 17.19~km and contains three main signalized intersections with through and turning movements. Road grade is set to zero throughout the corridor. In the experiments, each selected green phase is retained for $\Delta T_s=15$~s, and a 3-s yellow interval is inserted when the phase changes. The vehicle trajectory problem is solved every $\Delta t=1$~s with $H=15$, corresponding to a 15-s prediction horizon. The main simulation and controller parameters are reported in Appendix~\ref{app:parameter_settings} (Table~\ref{tab:parameter_settings}).

Mainline demand first increases and then decreases over the simulation horizon, producing a recurrent peak, while side-street demands remain constant. The fleet contains both petrol vehicles (PVs) and electric vehicles (EVs). CVs provide motion states and intended turning movements to the infrastructure, and the CAV subset additionally executes MP-adaptive ecological driving. Throughout the experiments, the reported CV penetration includes CAVs. All vehicles not controlled by the proposed trajectory optimizer, including non-connected vehicles and manually driven CVs, follow SUMO's default Krauss car-following model \citep{krauss1998microscopic}.

\subsubsection{Benchmark methods}\label{sssec:benchmark_methods}

Four control configurations are compared:
\begin{itemize}
    \item \textbf{Queue-based max-pressure (Q-MP) \citep{varaiya2013max}:} conventional MP control using queue-based movement states. All vehicles follow the default Krauss model.
    \item \textbf{Travel-time-based max-pressure (TT-MP) \citep{mercader2020max}:} MP control using accumulated travel-time states. All vehicles follow the default Krauss model.
    \item \textbf{Eco-MP:} the proposed signal controller using positive propulsion work as the movement state, with MP-adaptive ecological driving disabled.
    \item \textbf{Eco-MP with ecological driving (Eco-MP+ED):} the complete framework, combining Eco-MP signal control with MP-adaptive ecological driving for CAVs.
\end{itemize}
The comparison among Q-MP, TT-MP, and Eco-MP assesses the effect of the movement state used to construct pressure. The comparison between Eco-MP and Eco-MP+ED quantifies the additional contribution of vehicle-level trajectory control.

\subsubsection{Measures of effectiveness}\label{sssec:moe}

The evaluation uses four groups of indicators:
\begin{itemize}
    \item \textbf{Ecological performance:} cumulative positive propulsion work from Eqs.~\eqref{eq:positive_power}--\eqref{eq:positive_work}, reported in MJ/veh; PV fuel consumption from the Virginia Tech Comprehensive Power-Based Fuel Consumption Model, reported in L/100~km \citep{rakha2011virginia}; and EV electricity consumption from the corresponding electric-vehicle energy model, reported in kWh/100~km \citep{fiori2016power}. PV fuel use and EV electricity use are evaluated with their respective powertrain models, while the signal controller uses the common positive propulsion work defined in Section~\ref{sec:vehicle_power_work}.
    \item \textbf{Mobility performance:} vehicle delay relative to free-flow travel time and cumulative stopping time, defined as the total time for which vehicle speed remains below the same threshold $v^{\mathrm q}$ used in queue detection. In the microscopic simulation, this threshold follows the queueing-speed criterion used by SUMO.
    \item \textbf{Discharge performance:} stop-line time headway, calculated between consecutive vehicle crossings and averaged over observed departures.
    \item \textbf{Network operation:} the time-varying numbers of stopped vehicles, unserved vehicles still traveling or waiting to enter, and spillover vehicles unable to enter the modeled network.
\end{itemize}

Vehicle-level indicators are presented as distributions to show both central tendency and upper-tail behavior, while network-level indicators are reported as time series to capture congestion build-up and dissipation.

\subsection{Controller-comparison results}\label{ssec:comparison_results}

The base setting contains equal shares of PVs and EVs. Overall CV penetration is 50\%, and CAVs account for 30\% of the connected fleet, corresponding to 15\% of all vehicles. The demand profile and all other vehicle, signal, and trajectory-control parameters are identical across the four control configurations.

\subsubsection{Ecological performance}\label{sssec:ecological_results}

\begin{figure}[pos=t]
    \centering
    \includegraphics[width=\linewidth]{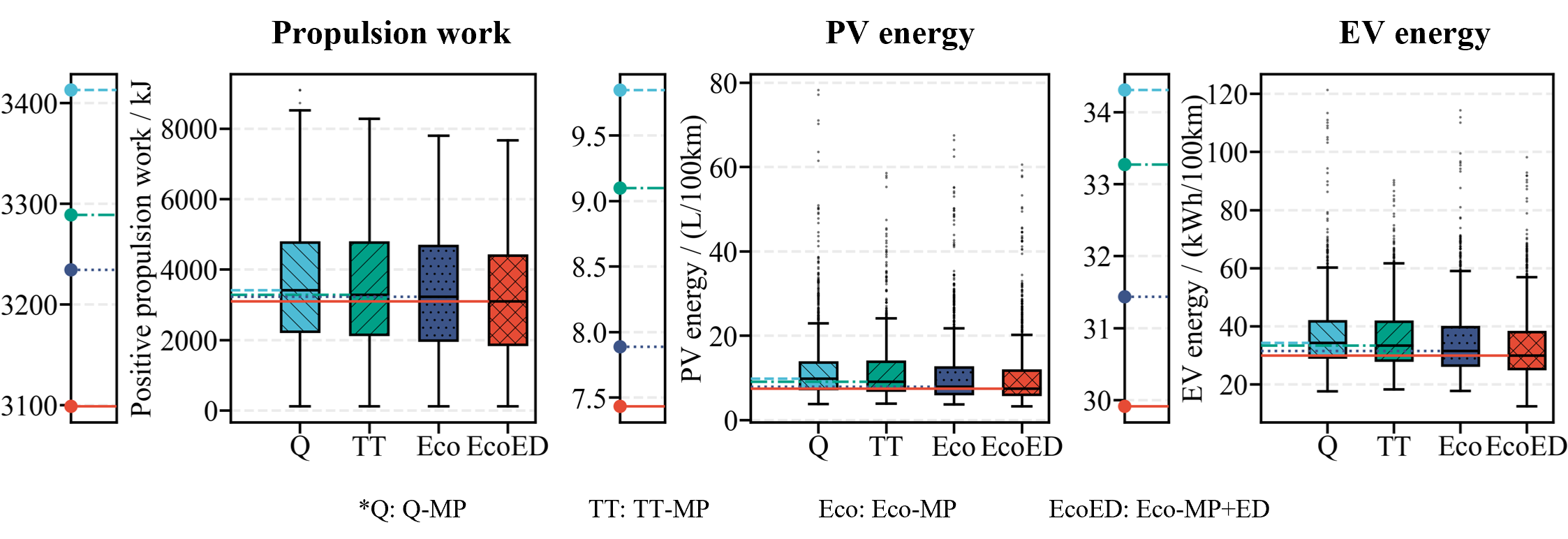}
    \caption{Vehicle-level ecological performance under different control configurations.}
    \label{fig:controller_energy}
\end{figure}

Among the signal-only controllers, Eco-MP achieves the lowest mean cumulative positive propulsion work in Fig.~\ref{fig:controller_energy}: 3.23~MJ/veh, compared with 3.41~MJ/veh under Q-MP and 3.29~MJ/veh under TT-MP. These values correspond to reductions of 5.2\% and 1.7\%, respectively. Because the three cases use the same traffic demand and car-following model and differ primarily in the movement state used to construct phase pressure, this comparison assesses the effect of ecological state design under otherwise common modeled conditions. A movement accumulates a larger Eco-MP state when more vehicles experience queueing and when those vehicles continue to accumulate positive propulsion work through stopping, restarting, and acceleration. Serving a movement with a large upstream ecological burden therefore removes vehicles carrying accumulated work and can also shorten the period over which additional stop-and-go work continues to accrue.

The powertrain-specific indicators follow the same ordering. Relative to Q-MP, Eco-MP reduces mean PV fuel consumption by 19.9\% and mean EV electricity consumption by 8.4\%; relative to TT-MP, the reductions are 13.4\% and 5.5\%, respectively. The larger relative reduction for PVs is consistent with the powertrain-specific evaluation models: EVs can recover part of the braking energy, whereas petrol vehicles have no corresponding regenerative-energy recovery. The consistent ordering of positive propulsion work and both powertrain-specific energy measures supports the use of positive propulsion work as a shared ecological state for mixed PV--EV traffic.

MP-adaptive ecological driving provides a second source of improvement. Relative to Eco-MP, Eco-MP+ED reduces mean positive propulsion work by 4.2\%, PV fuel consumption by 5.7\%, and EV electricity consumption by 4.8\%. The two layers act through complementary mechanisms: Eco-MP determines which movement receives service according to accumulated ecological pressure, while the vehicle controller shapes how each CAV approaches the resulting service opportunity. The acceleration penalty discourages abrupt speed corrections, and the signal constraints allow deceleration to begin before a predicted red interval becomes binding. The additional reductions are therefore consistent with smoother vehicle motion acting together with ecological phase selection.

\subsubsection{Trajectory-level mechanism}\label{sssec:trajectory_mechanism}

\begin{figure}[pos=t]
    \centering
    \includegraphics[width=\linewidth]{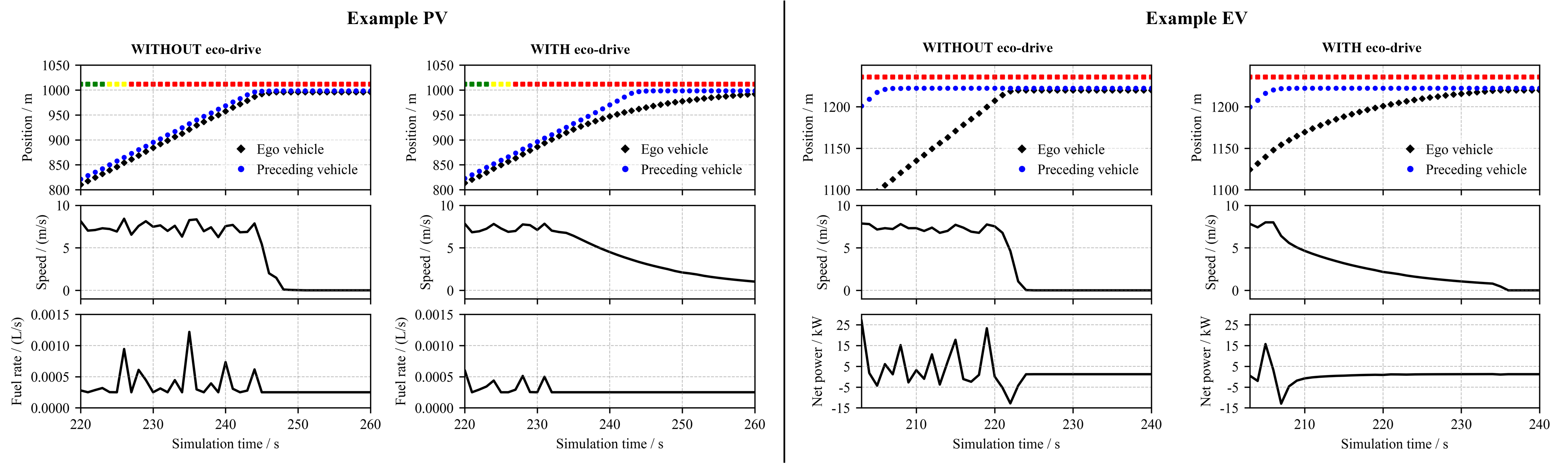}
    \caption{Representative PV and EV trajectories with and without MP-adaptive ecological driving.}
    \label{fig:representative_trajectories}
\end{figure}

The representative PV and EV trajectories in Fig.~\ref{fig:representative_trajectories} illustrate how MP-adaptive ecological driving changes the approach process. Without ecological driving, the example vehicles approach the intersection at relatively high speed and brake sharply as the red interval becomes binding, accompanied by repeated peaks in instantaneous power. With ecological driving, deceleration begins earlier and approach speed changes more gradually in response to the near-term signal prediction. This behavior follows directly from the finite-horizon formulation: the controller balances desired-speed tracking against acceleration effort while enforcing the predicted signal constraint, allowing an unavoidable stop to be anticipated before the vehicle reaches the stop line. The resulting trajectory reduces late braking, repeated acceleration, and unnecessary high-speed approach. The same qualitative pattern appears for the representative PV and EV, consistent with the powertrain-independent kinematic formulation and the aggregate reduction in positive propulsion work under Eco-MP+ED.

\subsubsection{Mobility and discharge performance}\label{sssec:mobility_results}

\begin{figure}[pos=t]
    \centering
    \includegraphics[width=\linewidth]{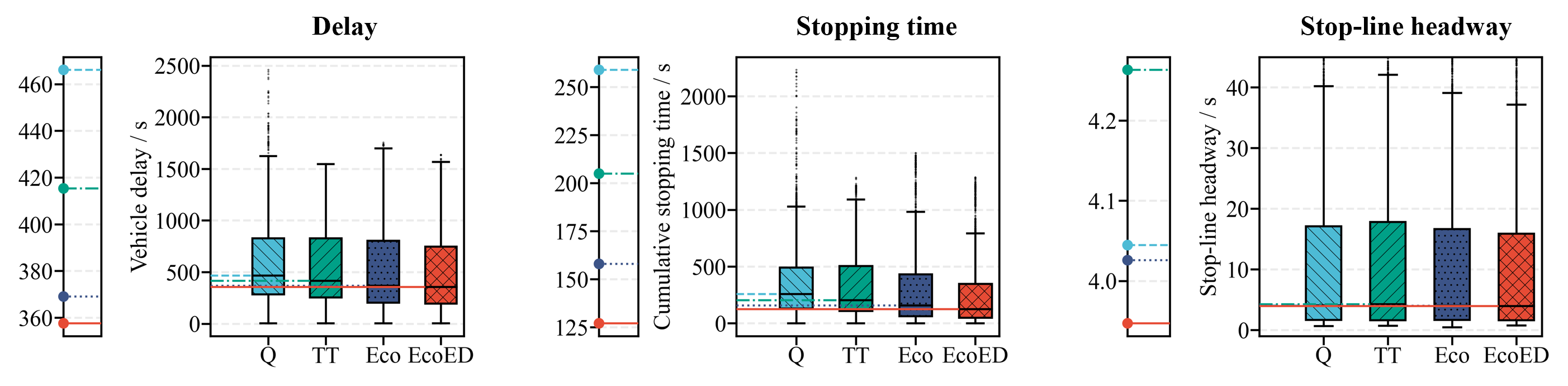}
    \caption{Vehicle-level mobility and discharge performance under different control configurations.}
    \label{fig:controller_mobility}
\end{figure}

The ecological gains are accompanied by lower delay and stopping time. As shown in Fig.~\ref{fig:controller_mobility}, Eco-MP reduces mean vehicle delay from 466~s under Q-MP and 416~s under TT-MP to 369~s, corresponding to reductions of 20.8\% and 11.2\%. Mean cumulative stopping time decreases from 259 and 205~s to 158~s, giving reductions of 39.0\% and 22.9\%, respectively. This joint improvement is consistent with the construction of ecological pressure. Under the tested demand, movements that accumulate substantial positive propulsion work typically contain vehicles that have experienced repeated stopping and restarting or prolonged queue discharge; continuing to defer these movements allows both congestion burden and ecological state to increase. Assigning greater service priority to such movements can therefore relieve stop-and-go traffic while reducing the propulsion burden represented by the controller.

Eco-MP+ED further reduces mean delay to 358~s and mean stopping time to 127~s. Relative to Eco-MP, these reductions are 3.1\% and 19.6\%, respectively. The larger change in stopping time reflects the role of the vehicle controller in shaping approach trajectories. Signal timing remains determined by Eco-MP, while the trajectory controller helps a CAV use a feasible green window or decelerate smoothly before a predicted red. This mechanism primarily reduces full stops and stop--restart cycles, which explains why stopping time changes more strongly than delay.

The stop-line headway results show a smaller but consistent operational improvement. Eco-MP+ED reduces mean headway from approximately 4.03~s under Eco-MP to 3.95~s, a reduction of about 2.0\%, corresponding to a slightly more compact departure process. With 15\% of all vehicles directly controlled, smoother CAV approaches can reduce delayed starts and improve use of available green intervals. This result is consistent with the service mechanism in Corollary~\ref{cor:service_region_noncontraction}: shorter effective discharge headways can increase realizable discharge, while any admissible-demand-region enlargement depends on whether those service gains enlarge the admissible-flow sets.

\subsubsection{Network-level operation}\label{sssec:network_results}

\begin{figure}[pos=t]
    \centering
    \includegraphics[width=0.8\linewidth]{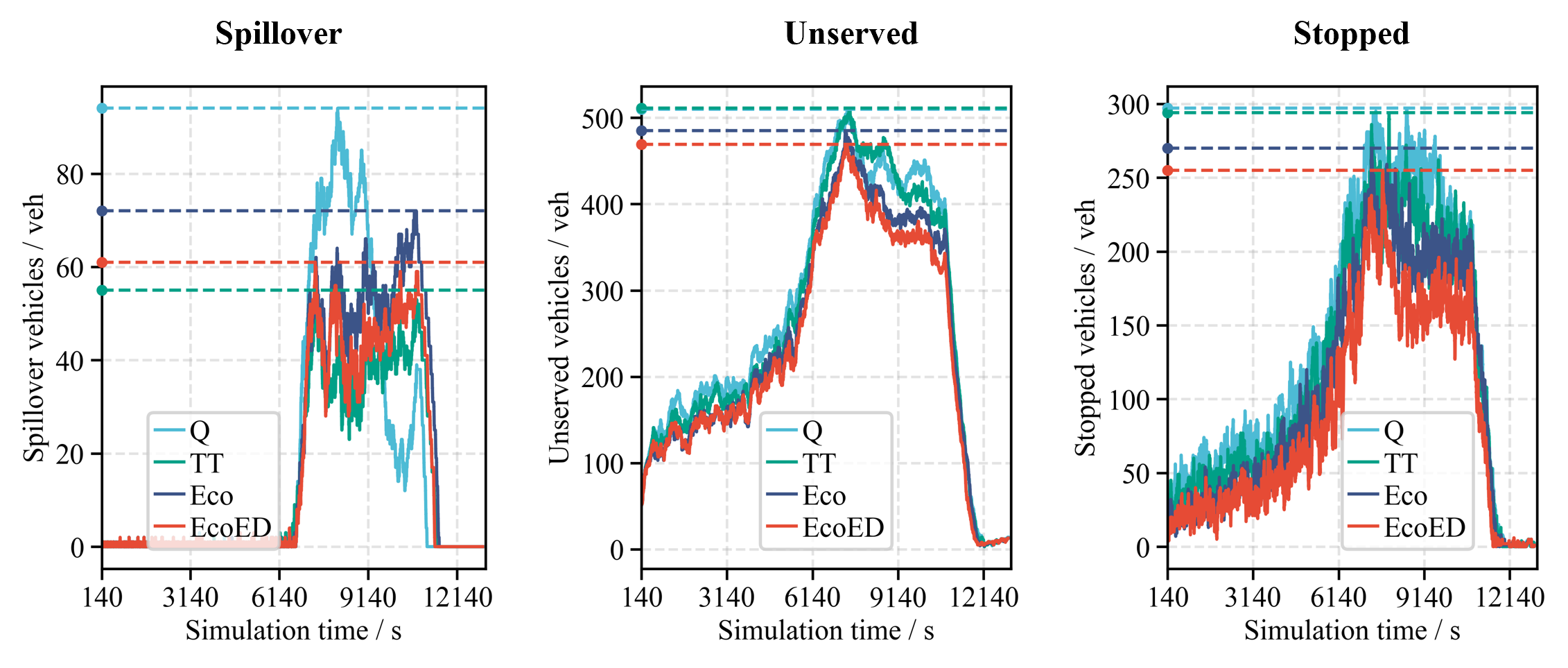}
    \caption{Network-level operational performance under different control configurations.}
    \label{fig:network_operation}
\end{figure}

The vehicle-level improvements are also reflected in the network accumulation shown in Fig.~\ref{fig:network_operation}. During the congested period, Eco-MP generally maintains fewer stopped and unserved vehicles than Q-MP and TT-MP. The peak number of unserved vehicles decreases from slightly above 500 under Q-MP to below 500 under Eco-MP, while the peak number of stopped vehicles decreases from about 300 to roughly 270. Eco-MP+ED further lowers these peaks to approximately 470 unserved vehicles and 255--260 stopped vehicles. The differences develop gradually over the peak period. Under the common demand profile, fewer unserved vehicles at the same simulation time indicate that a larger share of the same demand has already been processed. The lower accumulation therefore reflects faster congestion dissipation and greater demand processing over the peak period.

Boundary spillover shows a similar reduction relative to Q-MP. Its peak decreases from approximately 95 vehicles under Q-MP to about 70 under Eco-MP and 60 under Eco-MP+ED, while TT-MP reaches a slightly lower peak of approximately 55 vehicles. These indicators capture different aspects of congestion: TT-MP produces the smallest boundary queue in this experiment, whereas Eco-MP+ED produces the lowest peak numbers of stopped and unserved vehicles. Together with the headway result, this pattern indicates that the lower internal accumulation under Eco-MP+ED is not simply accompanied by a larger boundary queue.

\subsection{Sensitivity-analysis results}\label{ssec:sensitivity_results}

The sensitivity analysis examines how Eco-MP+ED responds to two deployment characteristics that can vary substantially in practice. The first experiment changes CV penetration, affecting both the traffic information available to Eco-MP and the share of vehicles capable of executing ecological driving. The second changes the PV--EV fleet composition while holding connectivity and control availability fixed, thereby isolating the effect of powertrain mix within a framework that uses a common ecological state and shared kinematic vehicle controller. Together, the experiments assess sensitivity to connectivity and applicability across mixed-powertrain traffic compositions.

\subsubsection{Sensitivity to CV penetration}\label{sssec:cv_sensitivity}

Eco-MP+ED is evaluated at CV penetration rates of 20\%, 50\%, and 80\%, with the PV share fixed at 50\%. CAVs remain 30\% of the connected fleet, so their shares of all vehicles are 6\%, 15\%, and 24\%, respectively. Increasing CV penetration therefore increases both the observation coverage available to Eco-MP and the proportion of vehicles receiving ecological-driving control.

\begin{figure}[pos=t]
    \centering
    \includegraphics[width=0.8\linewidth]{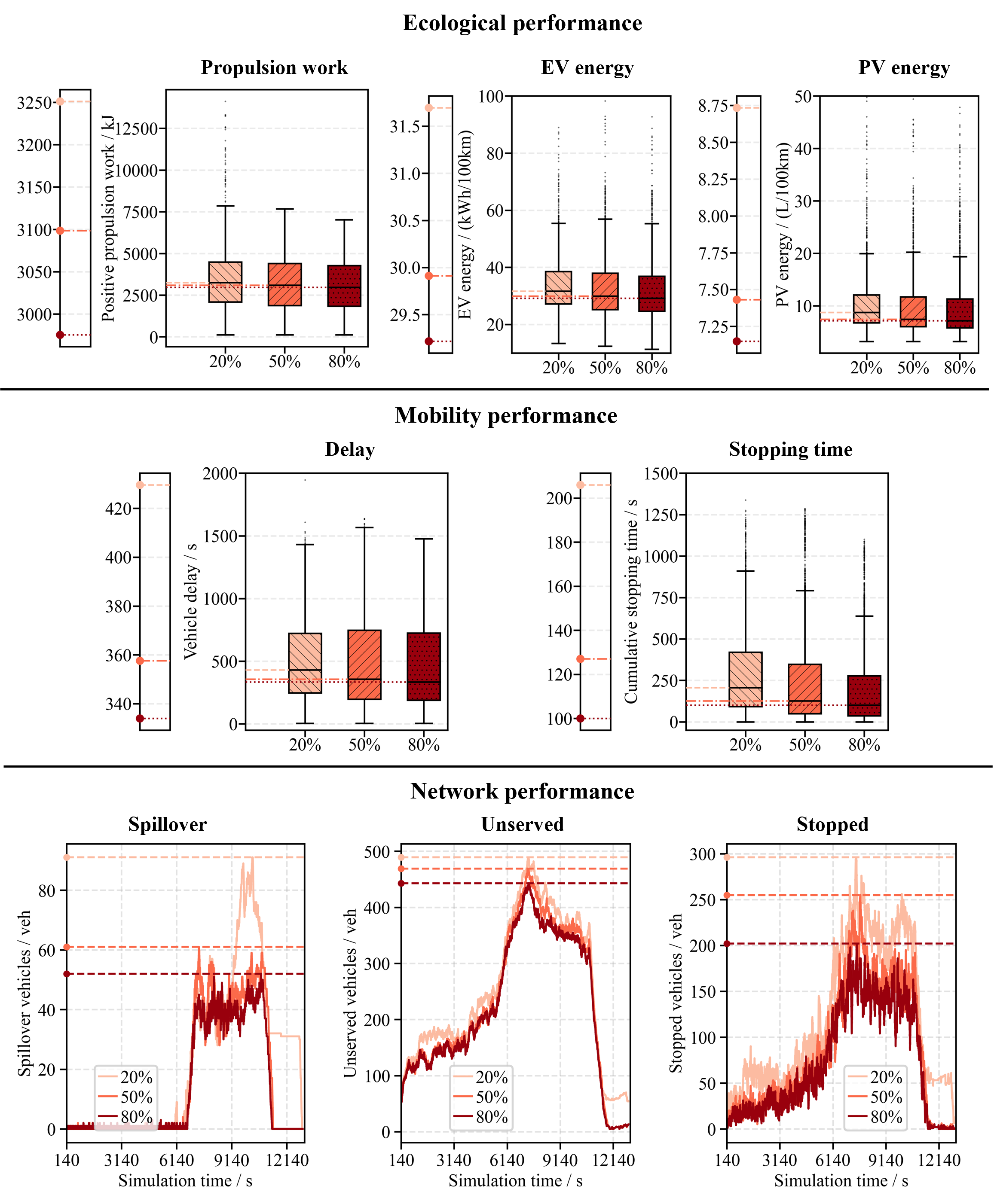}
    \caption{Performance under different CV penetration rates.}
    \label{fig:cv_performance}
\end{figure}

Performance improves systematically across the tested penetration levels in Fig.~\ref{fig:cv_performance}. From 20\% to 80\% CV penetration, mean positive propulsion work decreases from 3.25 to 2.98~MJ/veh, mean PV fuel consumption from 8.73 to 7.15~L/100~km, and mean EV electricity consumption from 31.70 to 29.22~kWh/100~km. Mean delay decreases from 430 to 334~s, while mean stopping time decreases from 206 to 100~s. These changes correspond to reductions of 8.3\%, 18.1\%, 7.8\%, 22.3\%, and 51.5\%, respectively.

The network time series follow the same ordering: higher penetration produces lower stopped, unserved, and spillover accumulation during the peak and faster recovery afterward. Two effects change simultaneously in this experiment. A larger connected sample provides broader coverage of the movement ecological states and intended turning movements used by Eco-MP, while the fixed 30\% CAV share within the connected fleet means that more vehicles also receive trajectory control. The pronounced reduction in stopping time is particularly consistent with the latter effect because the directly controlled share rises from 6\% to 24\%. For delay and stopping time, most of the improvement occurs between 20\% and 50\% penetration, with additional gains at 80\%, suggesting diminishing marginal benefits as both state coverage and controlled-vehicle presence increase. Under the tested joint deployment rule, substantial benefits are already observed at the intermediate penetration level.

\subsubsection{Sensitivity to fleet powertrain composition}\label{sssec:fleet_sensitivity}

To examine sensitivity to fleet electrification, the PV share is varied among 20\%, 50\%, and 80\% while Eco-MP+ED remains active. Overall CV penetration is fixed at 50\%, and CAVs remain 30\% of the connected fleet, corresponding to 15\% of all vehicles. The demand profile and all other parameters remain unchanged.

\begin{figure}[pos=t]
    \centering
    \includegraphics[width=0.8\linewidth]{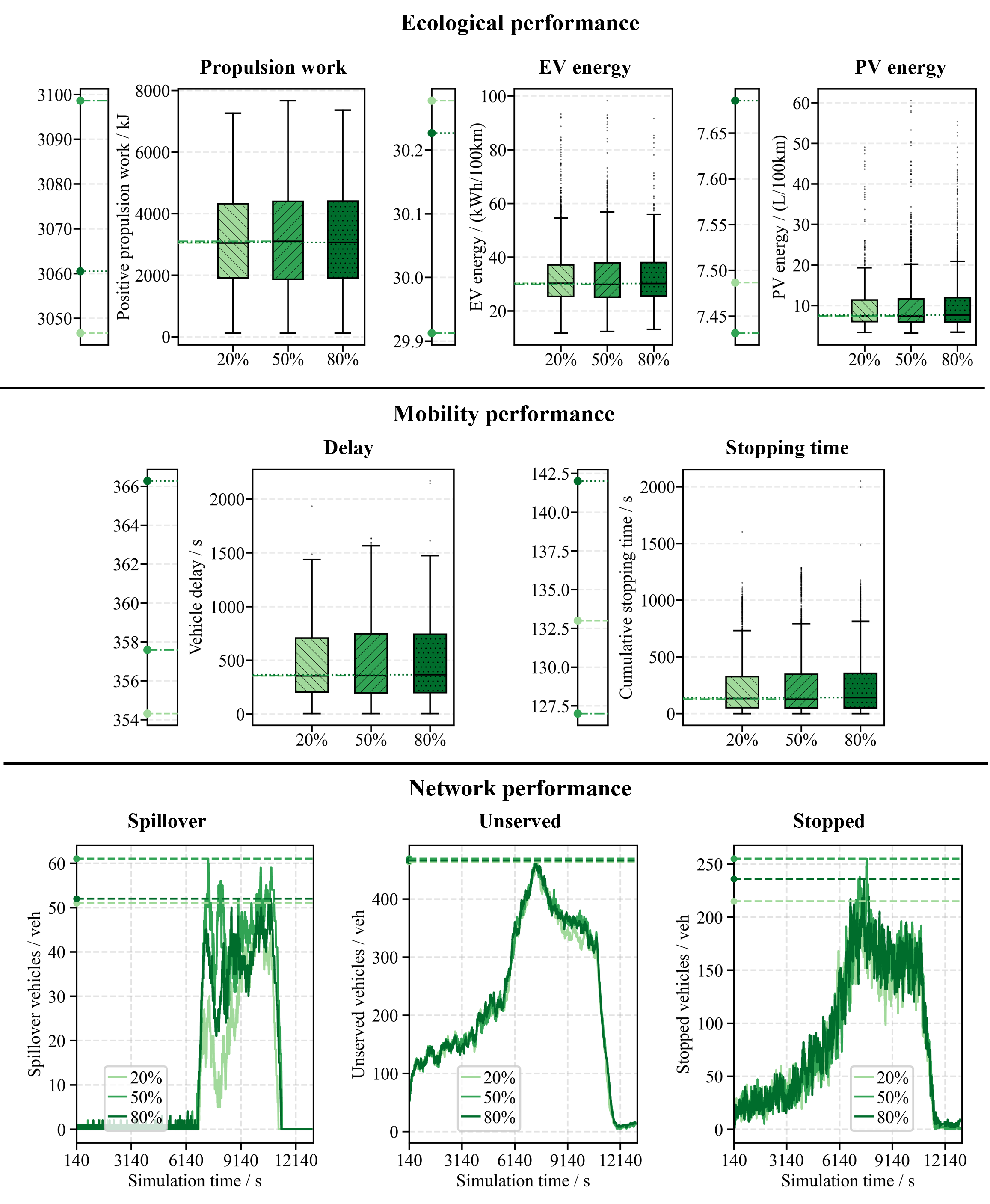}
    \caption{Performance under different PV shares.}
    \label{fig:pvshare}
\end{figure}

Performance varies only modestly across the three tested fleet compositions in Fig.~\ref{fig:pvshare}. Mean positive propulsion work remains within 3.06--3.10~MJ/veh, EV electricity consumption within approximately 30.2--30.3~kWh/100~km, and PV fuel consumption within 7.49--7.68~L/100~km. Mean delay ranges from 354 to 366~s and mean stopping time from 127 to 142~s. The corresponding network trajectories largely overlap and show no persistent ordering across the three PV shares.

This limited sensitivity indicates that the framework is relatively insensitive to fleet powertrain composition over the tested range. PVs and EVs contribute to the same ecological state defined by positive propulsion work, and the ecological-driving controller uses the same signal, car-following, and motion constraints for both powertrain types. The phase-selection logic responds to propulsion burden through this common state, while the vehicle controller reshapes motion through the same kinematic optimization. Changing the fleet mix therefore mainly changes how a given trajectory is translated into fuel consumption or electrical energy use, while traffic demand, connected-vehicle coverage, controlled-vehicle share, and signal-control structure remain fixed. This separation explains why varying the PV share from 20\% to 80\% has only a modest effect on the traffic and positive-propulsion-work indicators compared with changing the controller or CV penetration.

\subsection{Synthesis of experimental evidence}\label{ssec:experimental_synthesis}

The experiments separate the roles of the two control layers. The comparison among Q-MP, TT-MP, and Eco-MP shows that changing the movement state used for phase selection can reduce both propulsion burden and congestion. The Eco-MP--Eco-MP+ED comparison reveals an additional vehicle-level mechanism: rolling signal information produces smoother approaches and fewer full stops, together with a modest reduction in observed stop-line headway. At the network level, lower unserved accumulation and faster recovery show that these trajectory-level changes remain beneficial after propagating through the network.

These observations are consistent with the coupled analysis in Section~\ref{ssec:eco_driving_stability}. The trajectory-modified ecological state enters the same pressure-based stability framework, while the departure process determines the realizable services of the coupled model. Lower positive work, shorter stopping time, smaller headway, and lower network accumulation provide complementary evidence that the vehicle controller improves both trajectory-level and network-level operation in the tested corridor. The sensitivity experiments further show that the benefits increase as connectivity and direct vehicle control become more prevalent. The limited variation across substantially different PV--EV compositions also supports the intended mixed-powertrain design: both powertrains enter the same ecological state and use the same trajectory-control structure.

\section{Conclusion and Future Research}\label{sec:conclusion}

This paper developed Eco-MP to integrate ecological signal control and vehicle-level trajectory optimization within an MP framework for mixed-powertrain traffic. The signal layer uses a common movement state based on cumulative positive propulsion work after vehicles enter the detected queue region and constructs upstream--downstream ecological pressure for phase selection. The vehicle layer uses rolling signal information from the same feedback-driven controller and a powertrain-independent kinematic formulation to shape CAV trajectories. The resulting closed loop allows signal decisions and vehicle trajectories to adapt to one another through the evolving traffic and ecological states.

The theoretical analysis shows why this coupling remains compatible with the MP stability mechanism. A finite-storage Lyapunov function matched to the cumulative ecological state yields a control-dependent drift term equal to the ecological pressure--service objective maximized by Eco-MP. Under full-state information, Eco-MP strongly stabilizes admissible demands satisfying the stated strict supportability condition, and the same argument extends to the coupled system with trajectory-dependent realizable services. When ecological driving enlarges the admissible-flow sets, the coupled admissible demand region contains the signal-only region. The observed reduction in stop-line headway is consistent with this service-improvement mechanism and links the trajectory-level controller to the realizable services used in the coupled analysis.

The Amsterdam experiments support the two complementary control mechanisms. Relative to Q-MP, Eco-MP reduces mean positive propulsion work by 5.2\%, vehicle delay by 20.8\%, and stopping time by 39.0\%, with consistent reductions in PV fuel use and EV electricity use. These simultaneous gains reflect the fact that movements carrying a large propulsion burden also tend to contain vehicles that have experienced substantial stop-and-go congestion. Controlling only 15\% of the fleet with MP-adaptive ecological driving further reduces positive propulsion work by 4.2\% and stopping time by 19.6\%. The accompanying reduction in stop-line headway, lower unserved accumulation, and faster congestion recovery show that the trajectory-level improvements also carry through to network operation. Performance improves systematically as CV penetration increases and varies only modestly as the PV share changes from 20\% to 80\%, supporting the use of a common ecological state and trajectory-control structure across mixed-powertrain compositions.

Future research can extend the framework in three directions. First, ecological-state estimation and stability can be studied under incomplete, noisy, delayed, and privacy-sensitive CV observations. Second, uncertainty in rolling Eco-MP signal prediction can be incorporated into robust or stochastic trajectory optimization. Third, the vehicle layer can be extended from independent CAV control to coordinated multi-vehicle operation with explicit service-preserving and platoon-discharge constraints.

\appendix

\section{Simulation Parameter Settings}\label{app:parameter_settings}

Table~\ref{tab:parameter_settings} summarizes the main parameters used in the simulation experiments.

\begin{longtable}{p{0.22\textwidth}p{0.14\textwidth}p{0.42\textwidth}p{0.16\textwidth}}
\caption{Main parameter settings used in the simulation experiments.}\label{tab:parameter_settings}\\
\hline
Category & Symbol & Description & Value\\
\hline
\endfirsthead
\hline
Category & Symbol & Description & Value\\
\hline
\endhead
Simulation environment
& -- & PV share among all vehicles & 50\%\\
& -- & CV penetration & 50\%\\
& -- & CAV share among CVs & 30\%\\
& -- & Initial speed of vehicles entering the network & 8~m/s\\
\hline
MP signal control
& $\Delta T_s$ & Signal-control interval for the selected green & 15~s\\
& -- & Yellow interval & 3~s\\
& $c_m\Delta T_s$ & Saturation service per signal-control interval, through movement & 7~pcu/interval\\
& $c_m\Delta T_s$ & Saturation service per signal-control interval, left-turn movement & 6~pcu/interval\\
& $c_m\Delta T_s$ & Saturation service per signal-control interval, right-turn movement & 7~pcu/interval\\
\hline
Ecological state
& $g$ & Gravitational acceleration & 9.8~m/s$^2$\\
& $\mu_{\mathrm r}^k$ & Rolling-resistance coefficient & 0.015\\
& $C_D^k$ & Aerodynamic drag coefficient & 0.35\\
& $\rho_a$ & Air density & 1.225~kg/m$^3$\\
& $A_f^k$ & Vehicle frontal area & 2~m$^2$\\
& $\delta^k$ & Rotational-inertia coefficient & 1.1\\
& $M^k$ & Vehicle mass & 2000~kg\\
& $P_{\mathrm{aux}}^k$ & Auxiliary power demand & 2.5~kW\\
& $v^{\mathrm q}$ & Queue-identification speed threshold & 0.56~m/s\\
\hline
Ecological driving
& $q^{\mathrm{spd}}$ & Weight on speed-tracking error & 10\\
& $r^a$ & Weight on acceleration effort & 10\\
& $H$ & Prediction horizon & 15 steps\\
& $\Delta t$ & Vehicle-control and prediction time step & 1~s\\
& $v^{\mathrm{des}}$ & Desired speed & 16.67~m/s\\
& $a_{\max}$ & Maximum acceleration & 3~m/s$^2$\\
& $a_{\min}$ & Minimum acceleration & $-4$~m/s$^2$\\
& $v_{\max}$ & Maximum speed & 19.44~m/s\\
& $v_{\min}$ & Minimum speed & 0~m/s\\
& $v_{\mathrm{free}}^{\mathrm{IDM}}$ & IDM free-flow speed & 16.67~m/s\\
& $\delta_{\mathrm{IDM}}$ & IDM acceleration exponent & 4\\
& $d_0$ & IDM standstill gap & 10~m\\
& $T_h$ & IDM desired headway & 1~s\\
& $b$ & IDM comfortable-deceleration magnitude & 1.5~m/s$^2$\\
\hline
\end{longtable}

\section{Detailed Proofs for the Stability Analysis}\label{app:stability_proofs}

\subsection{Proof of Lemma~\ref{lem:eco_drift}}\label{app:proof_eco_drift}

Consider first a real movement $m=(i,o)$. Its vehicle count evolves according to traffic conservation,
\[
\dot z_m(t)
=
\sum_a r_{am}d_a(\bm s,t)-d_m(\bm s,t),
\]
and its ecological state satisfies $\dot\xi_m(t)=\alpha_m(t)-g_m(t)$. Between state-transition events,
\begin{align}
\frac{d}{dt}\!\left[\kappa_E z_m\xi_m\right]
&=
\kappa_E\xi_m
\left[
\sum_a r_{am}d_a(\bm s,t)-d_m(\bm s,t)
\right]
+
\kappa_Ez_m\alpha_m
-
\kappa_Ez_mg_m
\nonumber\\
&\leq
\kappa_E\xi_m
\left[
\sum_a r_{am}d_a(\bm s,t)-d_m(\bm s,t)
\right]
+
\kappa_Ez_m\alpha_m .
\label{eq:app_real_movement_drift}
\end{align}
Finite storage gives $z_m\leq\bar z_m$, while Eq.~\eqref{eq:eco_accumulation_bound} gives $\alpha_m\leq\bar z_m\bar P_m/L_i$. Hence $\kappa_Ez_m\alpha_m\leq\kappa_E\bar z_m^2\bar P_m/L_i$. Summing this bound over all real movements yields the finite constant $K_E$ in Lemma~\ref{lem:eco_drift}.

For a fictitious source movement $f$,
\[
\frac{d}{dt}\frac{z_f^2}{2}
=
z_f\lambda_f-z_fd_f(\bm s,t).
\]
The exogenous-arrival vector has nonzero entries only for fictitious source movements, so $\sum_fz_f\lambda_f=\bm w^\top\bm\lambda$. Summing Eq.~\eqref{eq:app_real_movement_drift} over real movements and the source-queue derivatives over fictitious movements gives
\begin{align}
\dot V(t)
&\leq
K_E
+
\bm w^\top(t)\bm\lambda(t)
-
\sum_a w_a(t)d_a(\bm s,t)
+
\sum_a d_a(\bm s,t)\sum_b r_{ab}w_b(t)
\nonumber\\
&=
K_E
+
\bm w^\top(t)\bm\lambda(t)
-
\sum_a
\left[
w_a(t)-\sum_b r_{ab}w_b(t)
\right]
d_a(\bm s,t)
\nonumber\\
&=
K_E
+
\bm w^\top(t)\bm\lambda(t)
-
\Psi(\bm s,t).
\label{eq:app_full_drift}
\end{align}
Conditioning on $\mathcal H_{t^-}$ and using $\mathbb E[\bm\lambda(t)\mid\mathcal H_{t^-}]=\bar{\bm\lambda}$ proves Eq.~\eqref{eq:eco_drift}.

Within the continuous-flow model, it remains to check the queue-memory updates represented as discrete state events rather than through $g_m$. When $\beta_m^k$ is first activated, $W_m^k$ starts from zero, so activation creates no positive jump in $\xi_m$. At vehicle discharge, let the pre-discharge state be $z_m^-=z$, $\xi_m^-=\xi$, and let the departing vehicle carry contribution $e\geq0$. The post-discharge state is $z_m^+=z-1$ and $\xi_m^+=\xi-e$, giving
\[
z_m^+\xi_m^+-z_m^-\xi_m^-
=
-\xi-(z-1)e
\leq0.
\]
Thus, queue-memory updates do not add a positive term to the drift bound.

\subsection{Proof of Theorem~\ref{thm:eco_mp_stability}}\label{app:proof_eco_stability}

Fix a mean demand $\bar{\bm\lambda}\in\Lambda^{\mathrm{int}}$ satisfying Definition~\ref{def:strict_supportability}. Then there exists a feasible reference policy $\bm s^{\circ}(\cdot)$ and constants $K^{\circ}<\infty$ and $\epsilon>0$ such that
\begin{equation}
K_E
+
\bm w^\top(t)\bar{\bm\lambda}
-
\Psi(\bm s^{\circ}(t),t)
\leq
K^{\circ}
-
\epsilon
\sum_f z_f(t).
\label{eq:app_reference_margin}
\end{equation}

For the current state, Eco-MP selects $\bm s^\star(t)$ to maximize
\[
\Psi(\bm s,t)
=
\sum_a\varpi_a(t)d_a(\bm s,t)
\]
over $\mathbb S$. Hence
\[
\Psi(\bm s^\star(t),t)
\geq
\Psi(\bm s^{\circ}(t),t).
\]
Combining this comparison with Lemma~\ref{lem:eco_drift} and Eq.~\eqref{eq:app_reference_margin} gives
\[
\mathbb E\!\left[
\left.
\dot V(t)
\right|
\mathcal H_{t^-}
\right]
\leq
K^{\circ}
-
\epsilon
\sum_fz_f(t).
\]
Integrating over $[0,T]$, taking expectations, and applying the tower property yields
\[
\mathbb E[V(T)]-\mathbb E[V(0)]
\leq
K^{\circ}T
-
\epsilon
\int_0^T
\mathbb E\!\left[
\sum_fz_f(t)
\right]dt.
\]
Since $V(T)\geq0$,
\[
\frac1T
\int_0^T
\mathbb E\!\left[
\sum_fz_f(t)
\right]dt
\leq
\frac{K^{\circ}}{\epsilon}
+
\frac{\mathbb E[V(0)]}{\epsilon T}.
\]
Taking $T\to\infty$ shows that the fictitious source queues have finite long-run average accumulation. For every real movement, Eq.~\eqref{eq:finite_real_storage} gives $z_m(t)\leq\bar z_m$. Strong network stability therefore follows from Definition~\ref{def:strong_stability}.

\subsection{Proof of Theorem~\ref{thm:coupled_stability}}\label{app:proof_coupled_stability}

Under MP-adaptive ecological driving, let $\tilde\alpha_m(t)$ denote the positive-propulsion-work accumulation rate associated with $\tilde\xi_m(t)$, and let $\tilde g_m(t)\geq0$ denote the rate at which this contribution leaves movement $m$. Bounded vehicle states and controls imply a finite movement-specific positive-power bound $\tilde{\bar P}_m$, so
\[
0\leq
\tilde\alpha_m(t)
\leq
\frac{\bar z_m\tilde{\bar P}_m}{L_i}.
\]
Consequently,
\[
\tilde K_E
=
\kappa_E\sum_m\frac{\bar z_m^2\tilde{\bar P}_m}{L_i}
<\infty.
\]

For a real movement,
\[
\dot z_m(t)
=
\sum_a r_{am}\tilde d_a(\bm s,t)
-
\tilde d_m(\bm s,t),
\]
and the coupled matched term satisfies
\begin{align*}
\frac{d}{dt}
\left[
\kappa_Ez_m\tilde\xi_m
\right]
&=
\kappa_E\tilde\xi_m
\left[
\sum_a r_{am}\tilde d_a(\bm s,t)
-
\tilde d_m(\bm s,t)
\right]
+
\kappa_Ez_m
(\tilde\alpha_m-\tilde g_m)\\
&\leq
\kappa_E\tilde\xi_m
\left[
\sum_a r_{am}\tilde d_a(\bm s,t)
-
\tilde d_m(\bm s,t)
\right]
+
\kappa_Ez_m\tilde\alpha_m.
\end{align*}
Adding the fictitious-source derivatives and repeating the rearrangement in Appendix~\ref{app:proof_eco_drift} gives
\begin{equation}
\mathbb E\!\left[
\left.
\dot{\tilde V}(t)
\right|
\mathcal H_{t^-},\bm s
\right]
\leq
\tilde K_E
+
\tilde{\bm w}^\top(t)\bar{\bm\lambda}
-
\tilde\Psi(\bm s,t).
\label{eq:app_coupled_drift}
\end{equation}
Queue-memory updates preserve this bound by the same argument as in Appendix~\ref{app:proof_eco_drift}.

Now fix a mean demand $\bar{\bm\lambda}\in\tilde\Lambda^{\mathrm{int}}$ satisfying Definition~\ref{def:coupled_strict_supportability}. Then there exists a feasible reference policy $\tilde{\bm s}^{\circ}(\cdot)$ and constants $\tilde K^{\circ}<\infty$ and $\tilde\epsilon>0$ such that
\[
\tilde K_E
+
\tilde{\bm w}^\top(t)\bar{\bm\lambda}
-
\tilde\Psi(\tilde{\bm s}^{\circ}(t),t)
\leq
\tilde K^{\circ}
-
\tilde\epsilon
\sum_fz_f(t).
\]
Because the coupled Eco-MP decision maximizes $\tilde\Psi$ over $\mathbb S$,
\[
\tilde\Psi(\bm s^\star(t),t)
\geq
\tilde\Psi(\tilde{\bm s}^{\circ}(t),t).
\]
Therefore,
\[
\mathbb E\!\left[
\left.
\dot{\tilde V}(t)
\right|
\mathcal H_{t^-}
\right]
\leq
\tilde K^{\circ}
-
\tilde\epsilon
\sum_fz_f(t).
\]
The integration argument of Theorem~\ref{thm:eco_mp_stability} then establishes finite long-run average accumulation of all fictitious source queues. Equation~\eqref{eq:finite_real_storage} bounds every real-movement count, so Definition~\ref{def:strong_stability} holds for the coupled system.

\subsection{Proof of Corollary~\ref{cor:service_region_noncontraction}}\label{app:proof_service_region_noncontraction}

Assume $\Omega(\bm s)\subseteq\tilde\Omega(\bm s)$ for every $\bm s\in\mathbb S$. Taking the union over feasible network signal schemes preserves set inclusion,
\[
\bigcup_{\bm s\in\mathbb S}\Omega(\bm s)
\subseteq
\bigcup_{\bm s\in\mathbb S}\tilde\Omega(\bm s).
\]
Taking convex hulls of both sides gives
\[
\Lambda
=
\operatorname{conv}
\left(
\bigcup_{\bm s\in\mathbb S}\Omega(\bm s)
\right)
\subseteq
\operatorname{conv}
\left(
\bigcup_{\bm s\in\mathbb S}\tilde\Omega(\bm s)
\right)
=
\tilde\Lambda.
\]
If the coupled admissible-flow sets contain at least one arrival vector outside $\Lambda$, the inclusion is strict.

\section*{Acknowledgment}
This work was partially supported by the National Key Research \& Development Program of China under Grant 2023YFB4301900; the National Natural Science Foundation of China under Grants 52572360, 52372319, 52302412, and 52372317; the Xiaomi Young Talents Program, and the Fundamental Research Funds for the Central Universities (22120230311). Hao Liu was supported by NSF Grant CMMI-2554235.

\section*{Declaration of generative AI and AI-assisted technologies in the manuscript preparation process}

During the preparation of this work, the authors used ChatGPT (OpenAI) to assist with language editing and to improve the clarity and organization of the manuscript. The authors subsequently reviewed and edited the content as needed and take full responsibility for the content of the published article.

\bibliographystyle{cas-model2-names}

\bibliography{Eco-MP}

\end{document}